\documentclass[aps,pra,superscriptaddress,reprint,floatfix,notitlepage,nobalancelastpage]{revtex4-2}
\usepackage[english]{babel}
\usepackage[utf8]{inputenc}
\usepackage{amsmath,amssymb}
\usepackage{graphicx}
\usepackage{url}
\usepackage{rotating}
\usepackage{float}
\usepackage{bbold}
\usepackage[colorinlistoftodos]{todonotes}
\usepackage[dvipsnames]{xcolor} 
\usepackage{empheq}
\usepackage{bm}
\usepackage[normalem]{ulem}
\usepackage{siunitx}
\usepackage{comment}
\usepackage{braket}
\usepackage[colorlinks=true,linkcolor=blue,urlcolor=blue,citecolor=blue]{hyperref} 
\makeatletter

\newcommand{\nb}{n_{\rm B}}
\newcommand{\be}{\begin{equation}}
\newcommand{\ee}{\end{equation}}
\newcommand{\bea}{\begin{eqnarray}}
\newcommand{\eea}{\end{eqnarray}}
\newcommand{\bse}{\begin{subequations}}
\newcommand{\ese}{\end{subequations}}

\begin{document}

\title{Analog Cosmology in a Quenched Bose-Einstein Condensate Bubble}

\author{Ruotong Zhai}
\affiliation{Department of Physics and Astronomy, Louisiana State University, Baton Rouge, LA 70803 USA}

\author{Brendan Rhyno}
\affiliation{Institut für Quantenoptik, Leibniz Universität Hannover, Welfengarten 1, 30167 Hannover, Germany}
\affiliation{Department of Physics, University of Illinois at Urbana-Champaign, Urbana, Illinois 61801, USA}

\author{Zain Mehdi}
\affiliation{Department of Quantum Science and Technology and Department of Theoretical and Fundamental Physics, The Australian National University, Canberra, ACT 2601, Australia}

\author{Naceur Gaaloul}
\affiliation{Institut für Quantenoptik, Leibniz Universität Hannover, Welfengarten 1, 30167 Hannover, Germany}

\author{Smitha Vishveshwara}
\affiliation{Department of Physics, University of Illinois at Urbana-Champaign, Urbana, Illinois 61801, USA}

\author{Daniel E. Sheehy}
\affiliation{Department of Physics and Astronomy, Louisiana State University, Baton Rouge, LA 70803 USA}

\date{October 14, 2026}

\begin{abstract}
 Recent interest and realization of ultracold bosons confined in shell-shaped, or bubble, traps have opened up several directions in the study of Bose-Einstein condensates (BECs) in intrinsically curved geometries.
Here, we explore the prospect of using thin spherical shell-shaped BECs as analog gravity simulators of an expanding or contracting curved universe.
We study an interacting BEC bubble subject to a temporal quench of the interaction strength.
Our theoretical approach, which uses a Bogoliubov treatment following a dimensional reduction of a spherical BEC to the low-energy excitations of a shell-shaped structure, demonstrates how this system serves as a laboratory analog for a positively curved universe having a time-varying scale factor.
We investigate spontaneous pair production triggered by quenches associated with exponentially decaying or growing interaction strengths, focusing on the emergence of oscillations in the structure factor---analogous to Sakharov oscillations in the Cosmic Microwave Background 
angular power spectrum.
We quantify the entanglement of the produced pairs via logarithmic negativity,
and show how this measure may be extracted from
the power spectrum. 
These findings establish a bridge between quantum information science and early-universe cosmology in a laboratory setting.
\end{abstract}

\maketitle

\section{Introduction}

Analog gravity, a field of physics that simulates cosmological phenomena, such as Hawking radiation~\cite{Hawking1974,Hawking1975} and particle production~\cite{Parker1966thesis,Parker1968}, using laboratory systems~\cite{Barcelo2011,Jacquet2020NextGen}, forms links across scales, from the subatomic to the astronomical. 
Associated experimental platforms include event horizons in nonlinear fiber optics~\cite{Philbin2008}, superfluids of light~\cite{Marino2008,Jacquet2020,Delhom2024}, Hawking radiation and the Unruh effect in intense laser fields~\cite{Schutzhold2006,Belgiorno2010}, superconducting quantum interference device arrays~\cite{Nation2009}, and even classical gravity wave analogs in water tanks~\cite{Schutzhold2002,Rousseaux2008}.

Ultracold atomic Bose-Einstein condensates (BECs), the focus here, offer a platform of wide-ranging interest and a highly controllable setting for analog gravity~\cite{Giacomelli2025} through tunable geometry and interatomic interactions. Due to their superfluid nature, the long-wavelength excitations of a BEC obey a linear dispersion relation, closely mimicking massless relativistic scalar fields. This feature has enabled the exploration of acoustic black holes~\cite{Visser1998,Garay2000,Garay2001,Recati2009,Lahav2010,Larne2012} and their corresponding quantum phenomena, including spontaneous Hawking radiation~\cite{Weinfurtner2011,Steinhauer2014,Steinhauer2016,Volovik2016,deNova2019,Kolobov2021}, the Unruh effect~\cite{Kosior2018,Hu2019,Gooding2020,Bhardwaj2023}, 
superradiance and ergoregion instabilities~\cite{Giacomelli2020,Giacomelli2021,Giacomelli2021b}, analog cosmic microwave background (CMB) reheating~\cite{Chatrchyan2021}, and Sakharov oscillations~\cite{Schmiedmayer2013,Hung2013,Song2026}.
Beyond black hole physics, BECs and superfluids serve as excellent analogous systems for simulating quantum fields in curved spacetime~\cite{Viermann2022,TolosaSimeon2022,SanchezKuntz2022,Tajik2023,Jarema2026}. In particular, dynamically tuning the interactions or background geometry allows one to simulate Friedmann-Lemaître-Robertson-Walker (FLRW) cosmologies with changing scale factor, and thus to investigate early-universe inflationary scenarios~\cite{Fedichev2003,Fedichev2004,Fischer2004,Jain2007,Llorente2019}.

Similar BEC experiments using  ring-shaped condensates have already been performed, offering a clean geometry for simulating closed 1D inflationary universes.
By physically expanding the ring, Eckel et al. successfully probed analog cosmological phenomena, such as phonon redshifts~\cite{Eckel2018,Eckel2021} and Hubble friction~\cite{Banik2022}.
Meanwhile, theoretical analyses of expanding toroidal BECs have predicted spontaneous and stimulated particle creation~\cite{Bhardwaj2021Inflationary} and quantified the bipartite entanglement of the resulting phonon pairs~\cite{Bhardwaj2024EntanglementBEC}.
\begin{figure*}
\vspace{.25cm}
    \centering
    \includegraphics[width=\linewidth]{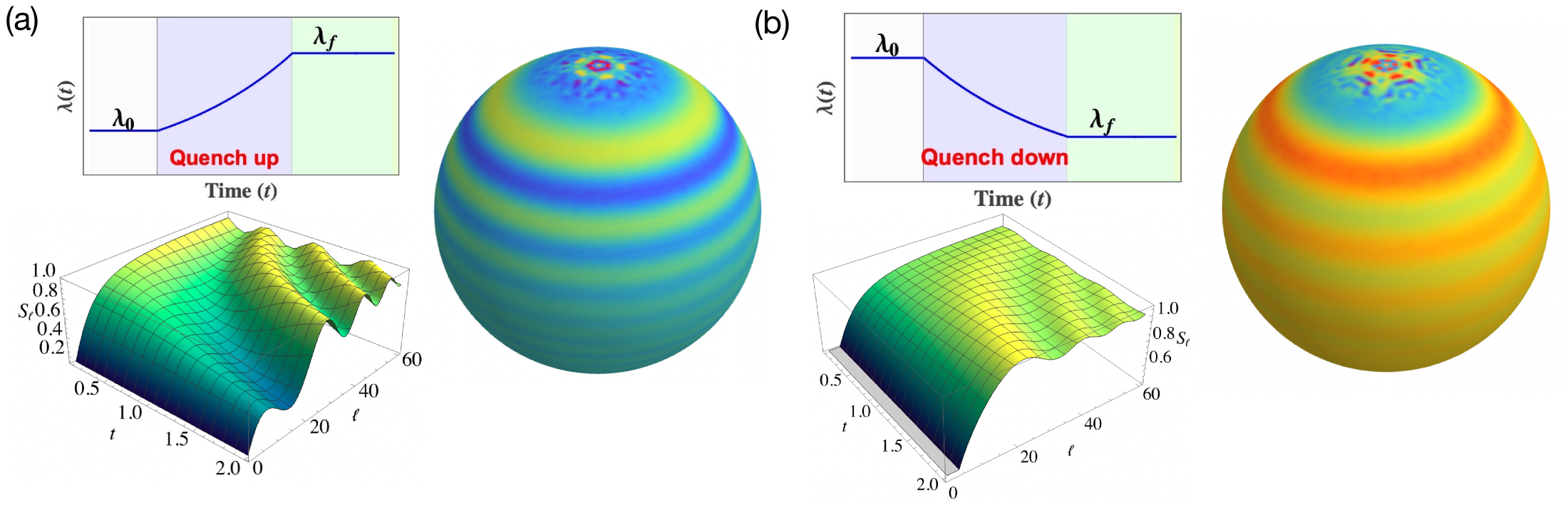}
    \caption{(Color online)
    Subfigure (a) shows how an increasing coupling
    ($\lambda(t)$) quench of a bubble BEC, shown in the upper-left plot, yields quantum particle production that is revealed in the time-dependent post-quench structure factor $S_\ell(t)$ (lower-left plot), with 
    $\ell$ the angular momentum quantum number, and in the post-quench boson density-density correlation function $\mathcal{C}_{\hat{\rho}\hat{\rho}}(\theta,\varphi,\theta',\varphi';t)$,
        defined in Eq.~(\ref{eq:densitycorrelator}),
        with $\theta$ and $\varphi$ the usual polar and azimuthal angles and
                $\theta' = 0$ (i.e. at the 
                \lq\lq north pole\rq\rq), with other details provided below Eq.~(\ref{densitycorreN}). Subfigure (b) shows the same plots for a decreasing-coupling quench, revealing qualitatively similar behavior (but reduced in magnitude).  }
    \label{fig:setup}
\end{figure*}
Our aim in this work is to study a thin spherical shell or \lq\lq bubble\rq\rq\ BEC as an analog gravity platform. A shell geometry would extend the analog universe from 1D to 2D. 
The creation of ultracold atomic BEC bubbles was originally proposed using radiofrequency (rf)-dressed potentials~\cite{Zobay2001,Zobay2004}.
Along this line, much theoretical work has been developed regarding the shell-shaped BECs, including  thermodynamics~\cite{Sun2018,Tononi2019,Bereta2019,Tononi2020,Rhyno2021,Rhyno2026}, vortices~\cite{Padavic2020,Bereta2021,Tononi2026}, collective modes~\cite{Lannert2007,Padavic2017}, phase transitions~\cite{Tononi2022}, and effects of the curved geometry~\cite{Tononi2023,Tononi2024,Tononi2025}..

However, realizing a single species shell-shaped BEC on Earth is challenging due to gravitational sag. This problem can be mitigated by going to a microgravity environment, such as a space station~\cite{Lundblad2019,Gaaloul2022,Elliott2023} or a drop tower facility~\cite{Zoest2010,Muntinga2013,Lotz2018}.
Within this approach, ultracold atomic bubbles were observed using the Cold Atom Lab (CAL) onboard the International Space Station~\cite{Carollo2022}.
On the other hand, Perrin et al. proposed a way to compensate for the gravitational sag on earth and observed expansions of BEC confined in a spherical-shell potential~\cite{Guo2022,Dubessy2025}.

Another direction for realizing a shell-shaped BEC in the presence of Earth's gravity involves using dual-~\cite{Wolf2022,Huang2025} or even triple-species~\cite{Ma2025,Ancilotto2026} atomic ensembles, appropriately tuning interatomic interactions to form the shell-shaped condensate.
Various experimental and theoretical studies have been conducted related to different aspects of dual-species shell-shaped BECs, including expansion dynamics~\cite{Jia2022}, the hollowing-out transition~\cite{Huang2025}, and the Lee-Huang-Yang correction beyond mean-field~\cite{Wen-Hao2025}.
Beyond bosonic systems, fermionic superfluid shells have also been theoretically investigated, including studies of the BCS-BEC crossover~\cite{He2022,Frigato2026}, dual-species shells~\cite{He2024,He2025}, and vortex physics~\cite{He2023,Sony2026}.

This Paper is organized as follows.  In Sec.~\ref{sec:BOG}, we develop the Bogoliubov theory for a two-dimensional (2D) spherical-shell BEC to simulate a 2D FLRW universe.
We demonstrate that dynamically tuning the atomic interaction strength—via a Feshbach resonance—can simulate either a ``Big Bang" (through a decreasing repulsive interaction strength) or a ``Big Crunch" (through an increasing repulsive interaction strength).
Within this framework, in Sec.~\ref{sec:spon} we calculate the resulting particle production driven by the analog expansion or contraction of the curved spacetime.
In Sec.~\ref{sec:SakharovOscillation}, we show that the calculated power spectrum of particle production exhibits a multi-peak structure characteristic of Sakharov oscillations~\cite{Sakharov1966,Grishchuk2012}, resembling the experimental observations by Hung et al.~\cite{Hung2013} who observed analog Sakharov oscillations in a quenched cold-atomic gas.
The setup and demonstration of the time evolution of the quantum fluctuations is shown in Fig.~\ref{fig:setup}.
In Sec.~\ref{sec:entanglement}, we quantify the entanglement of the produced particles using the logarithmic negativity measure~\cite{Peres:1996dw,Plenio:2005cwa,Simon:1999lfr} and show that the particle yield can be directly inferred from the time-dependence of the structure factor, mirroring the experimental approach of Chen et al.~\cite{Chen2021Quasiparticle} who measured pair production in a quenched atomic BEC. In Sec.~\ref{sec:concl}, we provide concluding remarks and discussion.  Appendix~\ref{appendix:dimensional_reduction}
discusses details of the dimensional reduction procedure, Appendix~\ref{appendix:Bogo_2-sphere} provides details of the Bogoliubov theory for the bubble BEC, and Appendix~\ref{appendix:realisticparameter} provides additional technical details including a discussion of realistic parameter values and our lengthy final result for the particle production amplitude.

\section{Bogoliubov Theory of Spherical-Shell BECs}
\label{sec:BOG}
We study a gas of  weakly-interacting bosons in three dimensions, with Hamiltonian:
\begin{align}
    \hat{H}(t)=\int_{\mathbb{R}^3} d^3 r\bigg[\hat\Psi^{\dagger}(\mathbf{r})\left(-\frac{\hbar^2}{2M}\nabla^2+V(\mathbf{r},t)-\mu\right)\hat\Psi(\mathbf{r})
    \nonumber\\
    +\frac{U_0(t)}{2}\hat\Psi^{\dagger}(\mathbf{r})\hat\Psi^{\dagger}(\mathbf{r})\hat\Psi(\mathbf{r})\hat\Psi(\mathbf{r})\bigg],
    \label{eq:H_3D}
\end{align}
where \(\hat\Psi\) and \(\hat\Psi^{\dagger}\) are bosonic annihilation and creation field operators, respectively.
Here, $M$ is the atomic mass, \(U_0(t)=4\pi\hbar^2a_s(t)/M\) represents a coarse-grained contact interaction with \(a_s(t)\) being the s-wave scattering length, \(V(\mathbf{r},t)\) is an external trapping potential that creates the spherical-shell geometry and $\mu$ is the chemical potential of the system that determines the equilibrium particle number $N$.  

To capture the essential features of a shell-shaped (or bubble)
BEC, we consider a spherically symmetric rf-dressed trapping potential of the form~\cite{Zobay2001,Zobay2004}:
\begin{equation}
    V(\mathbf{r},t)=\frac{1}{2}M\omega_0^2s_l^2\sqrt{[(|\mathbf{r}|/s_l)^2-\Delta(t)]^2+(2\Omega)^2},
    \label{eq:V_bubble}
\end{equation}
where $s_l=\sqrt{\hbar/2M\omega_0}$ is the oscillator length associated with the frequency \(\omega_0\), and \(\Delta\) and \(\Omega\) are dimensionless parameters that control the radius and width of the potential well \cite{Rhyno2026}. In the following, however, the specific form of the potential will not be of great importance, but rather we introduce an effective model that applies in the thin-shell limit.  

When the thickness is small compared to the radius of the shell, so that radial excitations are too energetically costly, we may assume that radial excitations freeze-out.
This is the thin-shell limit of a spherically symmetric bubble trap potential, and the effective low-energy theory is given by the following 2D Hamiltonian:
\begin{align}
    \hat{H}_{\mathrm{2D}}(t)=\int_{S^2} d\Omega\bigg[\hat\Psi^{\dagger}(\theta,\varphi) \left(\frac{L^2}{2MR(t)^2} - \mu_{\mathrm{2D}}(t) \right)
    \hat\Psi(\theta,\varphi)
    \nonumber\\
    +\frac{\lambda(t)}{2}\hat\Psi^{\dagger}(\theta,\varphi)\hat\Psi^{\dagger}(\theta,\varphi)\hat\Psi(\theta,\varphi)\hat\Psi(\theta,\varphi)\bigg],
    \label{eq:H_2D_eff}
\end{align}
where $\mu_{\mathrm{2D}}(t)$ is an effective
2D chemical potential that is renormalized
relative to the initial chemical potential~\cite{Tononi2025}.  (See Appendix~\ref{appendix:dimensional_reduction}.)

Here, $\theta$ and $\varphi$ are the usual polar and azimuthal angles on the sphere, respectively,  $d\Omega=\sin\theta d\theta d\varphi$ is the solid angle measure and 
 \(L^2\) is the usual angular momentum squared operator: 
\begin{equation}
L^2=-\hbar^2\left[\frac{1}{\sin\theta}\partial_{\theta}\sin\theta
\partial_{\theta}+\frac{1}{\sin^2\theta}\partial_{\varphi}^2\right]
 \equiv - \hbar^2\nabla^2,
 \label{eq:ellsquaredformula}
\end{equation}
with the last equality defining the Laplacian on a unit 2-sphere. 
The bosonic creation and annihilation operators
on the surface of the sphere satisfy the
commutation relations:
\begin{subequations}
\label{eq:2_sphere_bosonic_commutation_relations}
\begin{align}
    [\hat\Psi(\theta,\varphi),\hat\Psi^{\dagger}(\theta',\varphi')]
    &=
    \frac{1}{\sin \theta} \delta(\theta-\theta')
    \delta(\varphi-\varphi'),
    \\
    [\hat\Psi(\theta,\varphi),\hat\Psi(\theta',\varphi')]
    &=
    [\hat\Psi^{\dagger}(\theta,\varphi),\hat\Psi^{\dagger}(\theta',\varphi')]
    =
    0
    .
\end{align}
\end{subequations}
The effective 2D interaction strength $\lambda(t)$ depends on its 3D counterpart $U_0(t)$ as well as the spherical-shell radius $R(t)$ and thickness $\Delta R(t)$ via the approximate formula (see Appendix~\ref{appendix:dimensional_reduction}):
\begin{equation}
    \lambda(t) \sim \frac{1}{\sqrt{2\pi}}
    \frac{U_0(t)}{R(t)^2 \Delta R(t)}.
    \label{eq:coupling}
\end{equation}
This result tells us that the effective 2D interaction strength can be experimentally controlled by adjusting the radius or thickness of the thin spherical shell or by tuning the s-wave scattering length.

The dimensional reduction is justified when radial excitations remain frozen out. This requires two conditions: first, the temperature must be much smaller than the energy gap to higher radial modes, and second, the parameters in the Hamiltonian must vary slowly compared with the characteristic radial frequency. For a bubble trap in the large-detuning limit, the first condition can be written as
$k_{\mathrm{B}}T \ll \hbar\omega_{\mathrm{sh}}$, where
$\omega_{\mathrm{sh}} \equiv \omega_0\sqrt{\Delta/\Omega}$ is the characteristic radial trapping frequency. The corresponding shell thickness is approximately
$\Delta R \sim \sqrt{\hbar/(M\omega_{\mathrm{sh}})}$.
Combining these relations gives
\begin{equation}
    \Delta R \ll \sqrt{\frac{\hbar^2}{M k_{\mathrm{B}}T}},
\end{equation}
showing that the shell thickness must be much smaller than the thermal de Broglie wavelength (up to a numerical factor). Together with sufficiently slow temporal variations of the Hamiltonian parameters, this ensures that higher radial modes are not appreciably populated and that the dynamics remain effectively two-dimensional.

In the following, we assume the radius and shell thickness are constant (i.e. $R(t) = R$ and $\Delta R(t) = \Delta R$) but that $U_0(t)$ is time-dependent (controlled, for example, by a Feshbach resonance) leading to the time dependence in $\lambda(t)$. 

Our next task is to study the low-energy dynamics described by Eq.~(\ref{eq:H_2D_eff}).
In particular, we are interested in fluctuations on top of a spatially isotropic background BEC. 
This motivates us to write the bosonic field in terms of density and phase fluctuations using the Madelung representation:
\begin{equation}
\hat\Psi(\theta,\varphi)=\sqrt{\rho_0+\hat{\rho}(\theta,\varphi)}\; e^{i(\phi_0+\hat{\phi}(\theta,\varphi))},
    \label{Madelung}
\end{equation}
where $\rho_0$ and $\phi_0$ are the density and phase of the condensate.
For simplicity, we have neglected the time-dependence of the condensate here. However, a more careful analysis that takes the condensate dynamics into account leads to the same linearized Heisenberg equations of motion for the fluctuations (see Appendix~\ref{appendix:Bogo_2-sphere}).
The density and phase fluctuations on the surface of the sphere, $\hat{\rho}(\theta,\varphi)$ and $\hat{\phi}(\theta,\varphi)$, satisfy the canonical commutation relations:
\begin{subequations}
\label{eq:CCR_rho_phi}
\begin{align}
    [ \hat{\rho}(\theta,\varphi),\hat{\phi}(\theta',\varphi')]&=i\frac{1}{\sin\theta}\delta(\theta-\theta')\delta(\varphi-\varphi'),
    \label{commutatorphirho}
    \\
    [ \hat{\rho}(\theta,\varphi),\hat{\rho}(\theta',\varphi')]&=[ \hat{\phi}(\theta,\varphi),\hat{\phi}(\theta',\varphi')]=0.
\end{align}
\end{subequations}
Substituting Eq.~(\ref{Madelung}) into the effective 2D spherical-shell Hamiltonian, Eq.~(\ref{eq:H_2D_eff}), results in the Hamiltonian for density and phase fluctuations. Retaining only the quadratic terms yields
\begin{align}
    \hat{H}_{\mathrm{2D}}(t)
    = \int_{S^2} d\Omega
    &\Bigg[
    \frac{\hbar^2}{2MR^2} \rho_0
    \left( (\nabla \hat{\phi})^2 + \frac{1}{4\rho_0^2} (\nabla \hat{\rho})^2 \right)
    \nonumber\\
    &+
    \frac{1}{2} \lambda(t)\hat{\rho}^2
    \Bigg].
\end{align}
From this effective Hamiltonian one can derive the linearized Heisenberg equations of motion:
\begin{subequations}
\label{eq:rho_phi_fluc_EOM}
\begin{align}
    \partial_t\hat{\rho}&=-\frac{\hbar\rho_0}{MR^2}\nabla^2\hat{\phi}
    \label{eq:rho_fluc_EOM}
    ,
    \\
    -\frac{\hbar}{\lambda(t)} \partial_t\hat{\phi}&=\left(1-\frac{\hbar^2}{4MR^2\rho_0\lambda(t)}\nabla^2\right)\hat{\rho}
    \label{eq:phi_fluc_EOM}
    ,
\end{align}
\end{subequations}
for the density and phase fluctuations on the surface of the sphere in the presence of the
time-dependent coupling $\lambda(t)$.

\subsection{Analog Friedmann-Lema{\^i}tre-Robertson-Walker cosmology}

The analogy between BECs and relativistic scalar fields is well known and the basis of many studies of analog gravity \cite{Barcelo2011}.
Recently, this has been exploited in quasi-2D pancake-type geometries to study quantum fluctuations in analog expanding spacetimes \cite{TolosaSimeon2022,Viermann2022}.
Here, we see the manifestation of this on the 2-sphere.
In the long wavelength limit, or acoustic approximation, one neglects gradient terms in Eq.~(\ref{eq:phi_fluc_EOM})~\cite{Jain2007}.
This then allows one to use Eq.~(\ref{eq:phi_fluc_EOM}) to express the density fluctuations as a time-derivative of the phase fluctuations. Inserting this into Eq.~(\ref{eq:rho_fluc_EOM}) allows one to show that the phase fluctuations obey a massless Klein-Gordon equation in a curved background spacetime:
\begin{equation}
    0 = \square \hat\phi,
    \label{eq:KG}
\end{equation}
where
$
    \square \equiv \frac{1}{\sqrt{|g|}} \partial_\mu (\sqrt{|g|} g^{\mu\nu} \partial_\nu)
$
is the d'Alembertian operator in the curved spacetime, with $\mu = 0, 1, 2$ corresponding to $t, \theta, \varphi$ respectively, and the effective metric is:
\begin{align}
    g_{\mu\nu}(\theta,\varphi,t) &\equiv \text{const.} \left(\frac{\rho_0}{M R^2 c_s(t)}\right)^2
    \nonumber\\
    &\times\begin{pmatrix}
    -c_s(t)^2 & 0 & 0\\
    0 & R^2 & 0 \\
    0 & 0 & R^2 \sin^2\theta
    \end{pmatrix}
    .
    \label{eq:metric}
\end{align}
Here we have defined the speed of sound as
\begin{align}
    c_s(t) \equiv \sqrt{ \lambda(t) \rho_0 / M }
    .
\end{align}
Under this protocol, one has that the effective metric acts as an analog of the FLRW metric (up to a conformal factor which can be set to $1$ by choice of constant prefactor) with analog scale factor $a(t) = 1 / c_s(t) = \sqrt{M / \lambda(t) \rho_0}$. 
This leads to the line element on the 2-sphere:
\begin{equation}
ds^2 = -dt^2 + a^2(t) \left( R^2 d\theta^2 + R^2 \sin^2\theta d\varphi^2 \right),
    \label{eqn:lineelement}
\end{equation}
which is equivalent to a 2D FLRW metric:
\begin{equation}
    ds^2=-dt^2+a^2(t)R^2\left(\frac{du^2}{1-\kappa u^2}+u^2d\varphi^2\right),
\end{equation}
for $\kappa=1$. They are related by a coordinate change $u=\sin\theta/\sqrt{\kappa}$~\cite{TolosaSimeon2022, SanchezKuntz2022}.

These steps demonstrate that in the long-wavelength regime, the quenched spherical-shell BEC is governed by an effective relativistic Klein-Gordon equation. Standard acoustic approximations neglect the higher-order spatial derivatives—the Laplacian term in Eq.~(\ref{eq:phi_fluc_EOM})—restricting the analysis to strictly linear phononic excitations. In what follows, we keep the Laplacian term in Eq.~(\ref{eq:rho_phi_fluc_EOM}) to capture the full Bogoliubov dispersion, including the quadratic excitation of higher energy phonons. This allows us to investigate cosmological particle creation beyond the hydrodynamic acoustic limit.

\subsection{Mode Equations}

For a system confined to a spherical geometry, the eigenfunctions of the Laplace operator are the spherical harmonics: \(-\nabla^2Y_{\ell m}=\ell(\ell+1)Y_{\ell m}\).

These functions form a complete orthogonal basis on the sphere and thus allow small fluctuations be decomposed into independent angular momentum channels labeled by the quantum numbers $\ell=0,1,\dots$ and $m=-\ell,\dots,+\ell$.

This motivates the following mode expansions for the time-dependent density and phase fluctuations:
\begin{subequations}
\label{modeexpansion}
\begin{equation}
   \hat{\rho}(\theta,\varphi,t)=\sum_{\ell m}\Big[Y_{\ell m}(\theta,\varphi)\Pi_{\ell}(t)\hat{b}_{\ell m}
   +Y_{\ell m}^*(\theta,\varphi)\Pi_{\ell}^*(t)\hat{b}_{\ell m}^{\dagger}\Big],
   \label{modedecompositionrho}
\end{equation}
\begin{equation}
    \hat{\phi}(\theta,\varphi,t)=\sum_{\ell m}\Big[Y_{\ell m}(\theta,\varphi)\Phi_{\ell}(t)\hat{b}_{\ell m}
    +Y_{\ell m}^*(\theta,\varphi)\Phi_{\ell}^*(t)\hat{b}_{\ell m}^{\dagger}\Big],
    \label{modedecompositionphi}
\end{equation}
\end{subequations}
where $\hat{b}_{\ell m}$ and $\hat{b}_{\ell m}^{\dagger}$ are bosonic annihilation and creation operators associated with the excitation mode $(\ell,m)$. 
Physically, $\hat{b}_{\ell m}$ removes one quantum of excitation from the $(\ell,m)$ mode, while $\hat{b}_{\ell m}^{\dagger}$ creates one, and they obey the canonical commutation relation:
\begin{align}
    [\hat{b}_{\ell m},\hat{b}_{\ell' m'}^{\dagger}] &= \delta_{\ell \ell'}\delta_{m m'}
    ,
    \\
    [\hat{b}_{\ell m},\hat{b}_{\ell' m'}] &= [\hat{b}_{\ell m}^{\dagger},\hat{b}_{\ell' m'}^{\dagger}] = 0
    .
\end{align}
Because $\hat{\phi}$ and $\hat{\rho}$ describe fluctuations around a nonzero mean background, the uniform mode with $\ell=0$ and $m=0$ simply corresponds to a trivial global phase shift or a constant density offset, and are excluded from the sum.

We can plug the mode expansions, Eq.~(\ref{modeexpansion}), into the Heisenberg equations of motion, Eq.~(\ref{eq:rho_phi_fluc_EOM}) to get coupled mode equations for \(\Pi_\ell\) and \(\Phi_\ell\):
\begin{subequations}
\begin{align}
    \partial_t \Pi_\ell(t)
    &=
    \frac{\hbar\rho_0\ell(\ell+1)}{MR^2} \Phi_\ell(t)
    ,
    \label{modeeqnphi}
    \\
    -\frac{\hbar}{\lambda(t)} \partial_t \Phi_\ell(t) &=
    D_\ell(t) \Pi_\ell(t)
    ,
    \label{modeeqnp}
\end{align}
\label{modeeqnphiden}
\end{subequations}
where
\begin{align}
D_\ell(t) &\equiv 1+\frac{\hbar^2 \ell(\ell+1)}{4M R^2 \rho_0 \lambda(t)},
\label{eq:D_l}
\end{align}
is an eigenvalue of the differential operator that appears on the right-hand-side of Eq.~(\ref{eq:phi_fluc_EOM}). 

To proceed, we may combine the two
first order differential equations into one second order differential equation, eliminating either $\Pi_\ell(t)$ or $\Phi_\ell(t)$.  In the former case, eliminating the density mode coefficients yields 
\begin{multline}
\partial_t^2\Phi_{\ell}(t)=  \left( 2 \frac{\dot{c_s}(t)}{c_s(t)} + \frac{\dot{D_\ell}(t)}{D_\ell(t)} \right) \partial_t\Phi_{\ell}(t)\\
- D_\ell(t) \ell (\ell+1) \left( \frac{c_s(t)}{R} \right)^2 \Phi_{\ell}(t),
\label{phasemode}
\end{multline}
where the dot $\dot{} \equiv \partial_t$ denotes a time derivative. Similar to previous works on ring-shaped and pancake-type geometries, a Mukhanov-Sasaki equation emerges to describe the collective dynamics of each mode~\cite{Bhardwaj2021Inflationary,TolosaSimeon2022}.

Alternatively, eliminating the phase mode coefficients leads to the density mode equation:
\begin{equation}
\partial_t^2\Pi_\ell(t)=-\frac{\lambda(t)}{\hbar}\frac{\hbar\rho_0\ell(\ell+1)}{MR^2}D_\ell(t)\Pi_\ell(t),
\label{densitymode}
\end{equation}
where here we used that $R(t) = R$ is held fixed in our quench protocols.  

This density-mode equation describes the evolution of density excitations in the BEC in the presence of time-dependent system parameters and remains valid out of equilibrium. Since density correlations are commonly accessible observables in experiments, we use the density-mode equation, rather than the corresponding phase-mode equation, for the subsequent analysis.

\section{Spontaneous Particle Production}
\label{sec:spon}
\begin{figure}
    \centering
    \includegraphics[width=\linewidth]{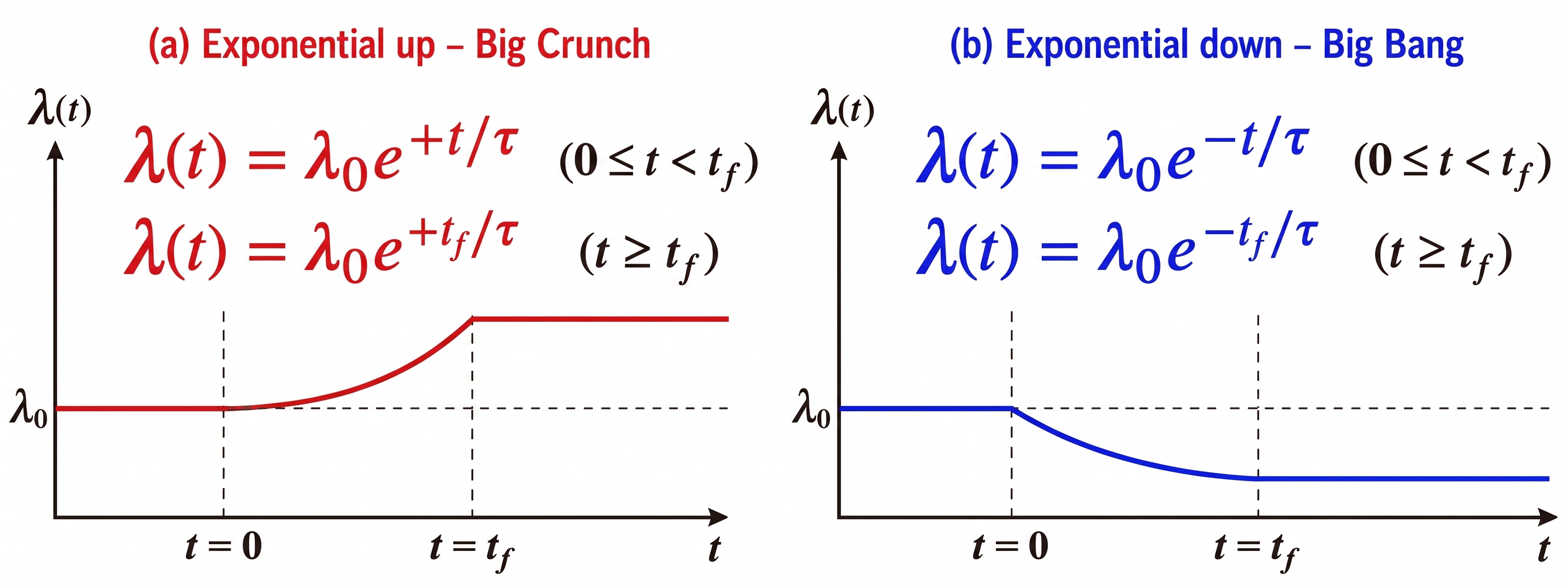}
    \caption{(Color online) The left and right panels illustrate two different coupling quench protocols. Since the analog scale factor is inversely proportional to the coupling strength, the increasing coupling quench (left panel) corresponds to a decreasing scale factor (or big crunch), while the decreasing coupling quench (right panel) corresponds to an increasing scale factor (or big bang).}
    \label{fig:quenchprofile}
\end{figure}

Equation~(\ref{eqn:lineelement}) 
implies that we can mimic an  analog expansion or contraction by controlling the effective scale factor  $a(t) \propto 1 / c_s(t) \propto 1/\sqrt{\lambda(t)}$.  In the following,
we therefore assume exponential 
control of the coupling, which goes through the following three-stage evolution:
\begin{equation}
    \lambda(t) =
\begin{cases}
  \lambda_0, & t \leq 0,  \hspace{.8cm}(\text{Regime 1})\\
  \lambda_0e^{\pm t/\tau},  & 0<t<t_f,\,(\text{Regime 2})\\
  \lambda_f \equiv \lambda_0e^{\pm t_f/\tau}, & t\geq t_f, \hspace{.7cm}
  (\text{Regime 3})
\end{cases}
\end{equation}
that we refer to as regimes $1$,$2$ and $3$ respectively. Fig.~\ref{fig:quenchprofile} shows the quench profile in the three regimes.
The exponentially increasing/decreasing coupling strength can replicate a contracting/expanding universe with Hubble constant $1/(2\tau)$.
In the following, we refer to the case with the plus sign as ``quench up" 
(corresponding to a contracting analog universe), and the case with the 
minus sign as ``quench down" (corresponding to an expanding analog universe).  
Next, we solve the mode equation in all three regimes.  We note that the 
mode equation solutions have a constant of motion, the Wronskian, that we 
choose to satisfy:
\begin{equation}
    \text{Wr}[\Pi_\ell,\Pi_\ell^*]=\frac{\hbar}{2\rho_0\varepsilon_\ell}(\Pi_\ell^*\dot{\Pi}_\ell-\dot{\Pi}_\ell^*\Pi_\ell)=-i,
    \label{Wronskian}
\end{equation}
where we introduced the single-particle energy
\begin{equation}
\varepsilon_\ell = \frac{\hbar^2}{2MR^2}\ell(\ell+1).
\end{equation}
In  regime 1, time $t\leq 0$, the coupling is constant and Eq.~(\ref{densitymode}) simplifies to: 
\begin{equation}
\partial_t^2\Pi_\ell^{(1)}(t)=-
\varepsilon_\ell \left( \varepsilon_\ell + 2\lambda_0\rho_0 \right)
\Pi_\ell^{(1)}(t).
\end{equation}
Solving this equation, we find a plane wave solution:
\begin{equation}
\Pi_{\ell}^{(1)}(t)=i\sqrt{\frac{\rho_0\varepsilon_\ell}{E_\ell^i}}e^{-iE_\ell^i t/\hbar},
    \label{inmodefunction}
\end{equation}
where \(E_\ell^i\) is  the initial Bogoliubov energy:
\begin{equation}
    E_\ell^i=\sqrt{\varepsilon_\ell \left( \varepsilon_\ell + 2\lambda_0\rho_0 \right)}.
    \label{Eq:ibo}
\end{equation}
The physical interpretation of 
Eq.~(\ref{inmodefunction}) is clear.  
Our initial system, a static spherical
shell BEC, can be regarded as a vacuum
(or thermal state) of
the mode operators $\hat{b}_{\ell m}$ with
$\hat{b}_{\ell m}^\dagger$ creating one
Bogoliubov quasiparticle with energy $E_{\ell}^i$.
When we plug Eq.~(\ref{inmodefunction}), 
 into the mode expansion for density fluctuations, Eq.~(\ref{modedecompositionrho}), we'll have a factor $e^{-iE_\ell^i t/\hbar}$ coming from $\Pi_{\ell}^{(1)}(t)$ multiplying the  annihilation operator $\hat{b}_{\ell m}$, and its complex conjugate 
$e^{iE_\ell^i t/\hbar}$ multiplying the
creation operator $\hat{b}_{\ell m}^\dagger$, so that the time-dependence of $\Pi_{\ell}^{(1)}(t)$ describes the
quantum evolution of these mode operators.  As we shall see, the 
coupling quench has the effect of mixing these plane wave solutions, which we interpret as particle creation.  

At time \(t=0\), the system enters
regime 2 with a time-dependent coupling.
In this regime, 
the mode equation takes the form:
\begin{equation}
    \partial_t^2\Pi_{\ell}^{(2)}(t)
    =
    -\left(
    \frac{2}{\hbar^2} \lambda_0 \rho_0 \varepsilon_\ell e^{\pm t/\tau}
    +
    \frac{\varepsilon_\ell^2}{\hbar^2}
    \right) \Pi_{\ell}^{(2)}(t).
\end{equation}
The mode equation in this regime is solved by imaginary indexed Bessel function: 
\begin{multline}
\Pi_{\ell}^{(2)}=A_{\ell}J_{-2i\sqrt{b_\ell}}(2\sqrt{a_\ell e^{\pm t/\tau}})\Gamma(1-2i\sqrt{b_\ell}) \\
    +B_{\ell}J_{2i\sqrt{b_\ell}}(2\sqrt{a_\ell e^{\pm t/\tau}})\Gamma(1+2i\sqrt{b_\ell}),
\end{multline}
where $A_\ell$ and $B_\ell$ are unknown coefficients (that will be fixed by matching to the regime 1 solution) and we defined the parameters $a_\ell$ and $b_\ell$:
\begin{subequations}
\begin{eqnarray}
a_\ell &\equiv&\frac{2\tau^2 \lambda_0 \rho_0 \varepsilon_\ell}{\hbar^2},
\\
    b_\ell &\equiv&\frac{\tau^2\varepsilon_\ell^2}{\hbar^2}.
\end{eqnarray}
\label{eqn:aandb}
\end{subequations}
At time $t_f$, the system enters 
regime $3$ in which the coupling
stops changing.  In this regime, the final coupling
value is  $\lambda(t_f)=\lambda_0 e^{\pm t_f/\tau}$ with the mode equation solution being a sum of complex exponentials:
\begin{equation}
\Pi_{\ell}^{(3)}(t)=i\sqrt{\frac{\rho_0\varepsilon_\ell}{E_\ell^f}}(\alpha_\ell e^{-iE_\ell^f t/\hbar}-\beta_\ell e^{iE_\ell^f t/\hbar}),
    \label{outmodefunction}
\end{equation}
where Eq.~(\ref{Wronskian}) implies 
$|\alpha_\ell|^2 -|\beta_\ell|^2 =1$ and
the final energy \(E_\ell^f\) is given by:
\begin{equation}
E_\ell^f=\sqrt{\varepsilon_\ell \left( \varepsilon_\ell + 2\lambda_0\rho_0 e^{\pm t_f/\tau} \right)}.
\label{eq:bogf}
\end{equation}
The coefficients \(A_\ell, B_\ell\) and \(\alpha_\ell, \beta_\ell\) can be obtained by matching the boundary condition and the derivative at the start and the end of the expansion:
\begin{align}
    &\Pi_{\ell}^{(1)}(0)=\Pi_{\ell}^{(2)}(0),\quad \dot{\Pi}_{\ell}^{(1)}(0)=\dot{\Pi}_{\ell}^{(2)}(0),\\
    &\Pi_{\ell}^{(2)}(t_f)=\Pi_{\ell}^{(3)}(t_f),\quad \dot{\Pi}_{\ell}^{(2)}(t_f)=\dot{\Pi}_{\ell}^{(3)}(t_f).
\end{align}
Since the details of this calculation are
straightforward, we only present our lengthy expressions for $\alpha_\ell$ and $\beta_\ell$ in Appendix~\ref{appendix:realisticparameter}. 

Physically, we interpret $|\beta_\ell|^2$ as the number of particles (or quanta) with quantum number $\ell$ produced by the quench.
To see this, note that
the form of the final
mode function in Eq.~(\ref{outmodefunction}) implies that
the coupling quench has converted our
initial plane wave solution
$\Pi_{\ell}^{(1)}(t)\propto {\rm e}^{-iE_\ell^it/\hbar}$
in to a sum of complex exponentials.

This furthermore implies the quench imparts a Bogoliubov transformation on the mode operators $\hat{b}_{\ell m}$. To see the latter, we plug Eq.~(\ref{outmodefunction}) into Eq.~(\ref{modedecompositionrho}) and rearrange terms so that all terms with factors  $\propto {\rm e}^{iE_\ell^f t/\hbar}$ (or 
$\propto {\rm e}^{-iE_\ell^f t/\hbar}$) 
multiply annihilation (or creation) operators.  This defines
\lq\lq final\rq\rq annihilation operators:
    \begin{equation}
        \hat{c}_{\ell m}=\alpha_\ell\hat{b}_{\ell m}+\beta_\ell^*(-1)^m\hat{b}_{\ell,-m}^{\dagger},
        \label{BogoliubovTransformation}
    \end{equation}
    that mix particle and holes of
    the original system.  

\begin{figure}[!]
    \centering
    \includegraphics[width=0.8\linewidth]{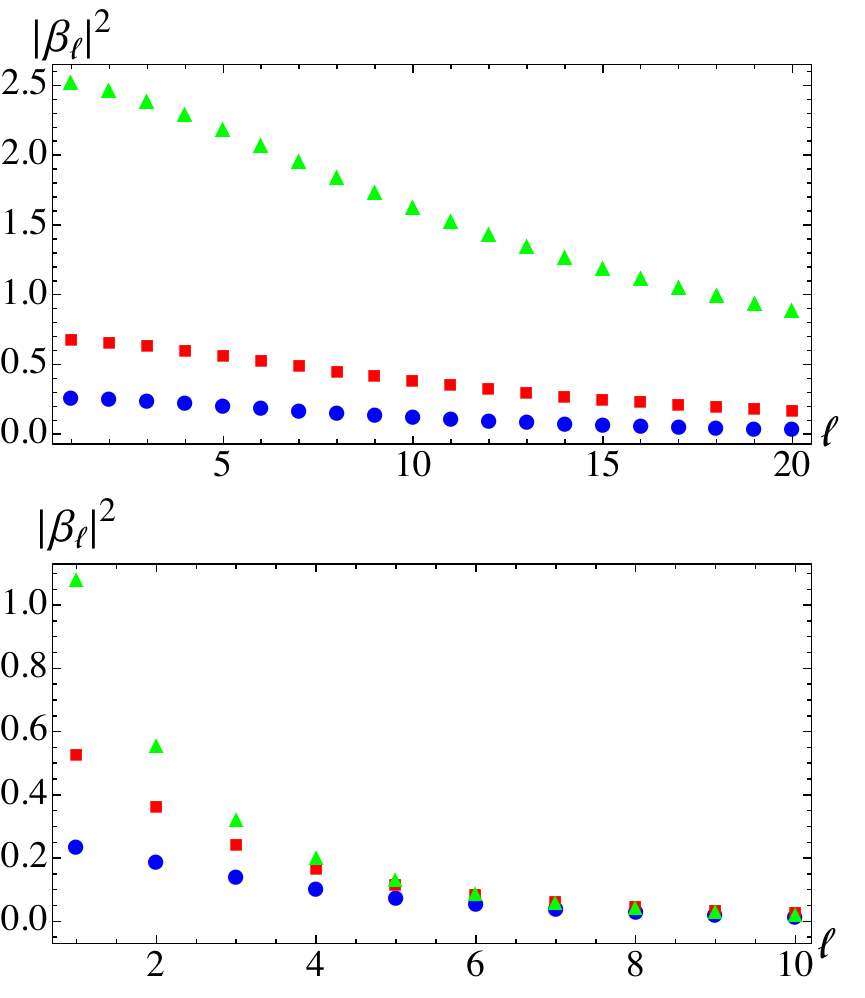}
    \caption{(Color online) The top panel shows the particle production vs. angular momentum 
    quantum number $\ell$ for
    a quench-up experiment 
    (i.e., an analog big crunch) and the bottom panel shows the same for a quench-down experiment (i.e., an analog big bang).
    In both
    panels, the  quench durations are $t_f=20\,\mathrm{ms}$ (blue circles), $t_f=30\,\mathrm{ms}$ (red squares), and $t_f=50\,\mathrm{ms}$ (green triangles).
    This can be viewed as a spectrum of particles or quanta created by the quench, and 
    in each case the  particle production  monotonically decreases with increasing $\ell$. For the parameters considered here, the quench-up protocol produces more such quanta than the corresponding quench-down protocol.  This is likely because increasing the coupling strength leads to a more robust BEC. 
    }
    \label{Betasquare}
\end{figure}

To understand the particle production picture of this process, recall that 
our initial system in region 1 has a 
well-defined vacuum $| 0_b \rangle$ that is annihilated
by the mode operators $\hat{b}_{\ell m}$.  Clearly, the expectation value of 
the number of $b$ particles  gives a vanishing result in this initial state
\begin{equation}
\langle 0_b | \hat{b}_{\ell m}^\dagger
\hat{b}_{\ell m}| 0_b \rangle = 0. 
\end{equation}
After the interaction quench, Eq.~(\ref{BogoliubovTransformation})
implies (using the Heisenberg picture), a 
nonzero number of \lq\lq c\rq\rq\ particles:
\begin{equation}
\langle 0_b | \hat{c}_{\ell m}^\dagger
\hat{c}_{\ell m}| 0_b \rangle = |\beta_{\ell}|^2.
\end{equation}
Typical results for $|\beta_{\ell}|^2$ are shown 
in Fig.~\ref{Betasquare}, showing a larger particle
production in the increasing-coupling quench 
(an analog big crunch) as compared to the decreasing-coupling quench (an analog big bang).
For both cases, the particle production is also
strongest at small $\ell$.  

One insight that can
explain both of these trends is that the particle
production process relies on the superfluid nature
of excitations in an interacting BEC.  At large
$\ell$, excitations lose their superfluid identity, as follows from the large-$\ell$ limit of
the Bogoliubov energies Eq.~(\ref{Eq:ibo}) 
or Eq.~(\ref{eq:bogf}), i.e., 
$E_\ell^{i/f} \propto \varepsilon_\ell$.  Thus, we
expect no particle production at large $\ell$.
Since a quench up (quench down) strengthens 
(weakens) superfluid correlations, we similarly expect a stronger
particle production in the quench up case. 
Our next task is to study 
experimental signatures of this 
particle production.

\section{Sakharov Oscillations}
\label{sec:SakharovOscillation}
We propose that one specific signature of particle production in a quenched spherical-shell BEC is a cold-atom analog of Sakharov oscillations. In cosmology, these oscillations arise from acoustic-wave interference in the primordial plasma and are observed as a series of harmonic peaks in the angular power spectrum of the Cosmic Microwave Background (CMB)~\cite{Grishchuk2012}. In cold-atom experiments, the static structure factor plays the role of this cosmological power spectrum.

Cold-atom experiments have already observed analog Sakharov oscillations, including the early work of Hung et al.~\cite{Hung2013}, who studied the evolution of density correlations in an atomic BEC after an interaction quench, observing time-dependent oscillations in the structure factor. More recently, Viermann et al.~\cite{Viermann2022} observed density fluctuation correlations in an analog curved-spacetime BEC that exhibited a time evolution consistent with analog Sakharov oscillations. In both experiments, the Sakharov oscillations manifested as a multi-peak structure in the structure factor (power spectrum) that compresses with time.

In the following, we consider whether the predicted particle production would show similar signatures in a quenched spherical-shell BEC. Following the earlier experiments, it is natural to consider the equal-time density-density correlation function, 
 defined on the surface of the sphere:
\begin{equation}
    \mathcal{C}_{\hat{\rho}\hat{\rho}}(\theta,\varphi,\theta',\varphi';t)=\braket{\hat{\rho}(\theta,\varphi,t)\hat{\rho}(\theta',\varphi',t)}
    \label{eq:densitycorrelator} 
\end{equation}
where the angle brackets represent
a thermal expectation value.
Here, the density fluctuation operators are given  by Eq.~(\ref{modedecompositionrho}), where for the mode functions $\Pi_\ell(t)$ we must use our results from 
the preceding section.  As stated above, 
we assume the initial system to be an interacting BEC at temperature $T$, so that the mode operator averages are:
\begin{align}
    & \braket{\hat{b}_{\ell m}\hat{b}_{\ell m}} =\braket{\hat{b}_{\ell m}^{\dagger}\hat{b}_{\ell m}^{\dagger}}=0,\\
    & \braket{\hat{b}_{\ell m}^{\dagger}\hat{b}_{\ell m}} =\nb(E_\ell^i),\\
    & \braket{\hat{b}_{\ell m}\hat{b}_{\ell m}^{\dagger}} =1+\nb(E_\ell^i),
\end{align}
where $\nb(E_\ell^i)=1/(e^{E_\ell^i/(k_{\mathrm{B}}T)} - 1)$ is the Bose-Einstein distribution function
with $T$ the initial system temperature. 

Before the quench, in the initial 
 interacting BEC state, 
the correlator $\mathcal{C}_{\hat{\rho}\hat{\rho}}$ is time-independent (as
expected for an equilibrium state).  
For this reason, we immediately
proceed to the post-quench regime, in 
which the mode-function solutions 
are given by Eq.~(\ref{outmodefunction}),
with the effect of the quench encoded in the particle production parameter $\beta_\ell$.  We find:
\begin{multline}
    \mathcal{C}_{\hat{\rho}\hat{\rho}}(\theta,\varphi,\theta',\varphi';t)
    =
    \sum_{\ell=1}^\infty \frac{2\ell+1}{4\pi}
    P_\ell(\cos L)
    \frac{\rho_0\varepsilon_\ell}{E_\ell^f}
    \\
\times     \Big( 
    1+2|\beta_\ell|^2-2\Re[\alpha_\ell^*\beta_\ell e^{2 i E_\ell^f t / \hbar}]\Big)[1+2\nb(E_\ell^i)],
    \label{densitycorreN}
\end{multline}
where the angle dependence enters entirely via  $L\equiv \cos^{-1}(\cos\theta\cos\theta'+\sin\theta\sin\theta'\cos(\varphi-\varphi'))$, the geodesic length on a 2D sphere.  In Fig.~\ref{fig:setup}, we show 
$\mathcal{C}_{\hat{\rho}\hat{\rho}}(\theta,\varphi,\theta',\varphi';t)$ for the
quench-up case, where we chose
$\theta' = 0$ (at the north pole) and varying $\theta$ and $\varphi$. 
With this choice of angle, $L$ is clearly independent of 
$\varphi$, yielding the concentric rings seen in this plot. In Fig.~\ref{fig:SakharovOscillation}, we chose time stamp $t =0\,\mathrm{s}$, $0.3\,\mathrm{s}$, $0.9\,\mathrm{s}$, and $1.2\,\mathrm{s}$, and other parameters  consistent with the CAL experiments as discussed in Appendix~\ref{appendix:realisticparameter}.
We emphasize that  $\mathcal{C}_{\hat{\rho}\hat{\rho}}$ is explicitly time-dependent due to the quench-induced particle production, and that this time dependence would vanish in the limit of zero particle production, $\beta_\ell \to 0$.  

A more convenient way to express the same information as in $\mathcal{C}_{\hat{\rho}\hat{\rho}}$ is to consider its 
 \lq\lq momentum space\rq\rq\
analog, obtained by performing a 
Fourier-Legendre transformation: 
\begin{equation}
    S_\ell(t)=\frac{2\ell+1}{2}\int_{-1}^1 P_\ell(\cos L)\mathcal{C}_{\hat{\rho}\hat{\rho}}(L,t)d(\cos L),
\end{equation}
where we have written the argument of 
$\mathcal{C}_{\hat{\rho}\hat{\rho}}$
in a slightly different way.  Using this, we obtain $S_\ell(t)$, the structure factor of the post-quench BEC:
\begin{figure*}[t]
    \centering
    \includegraphics[width=\linewidth]{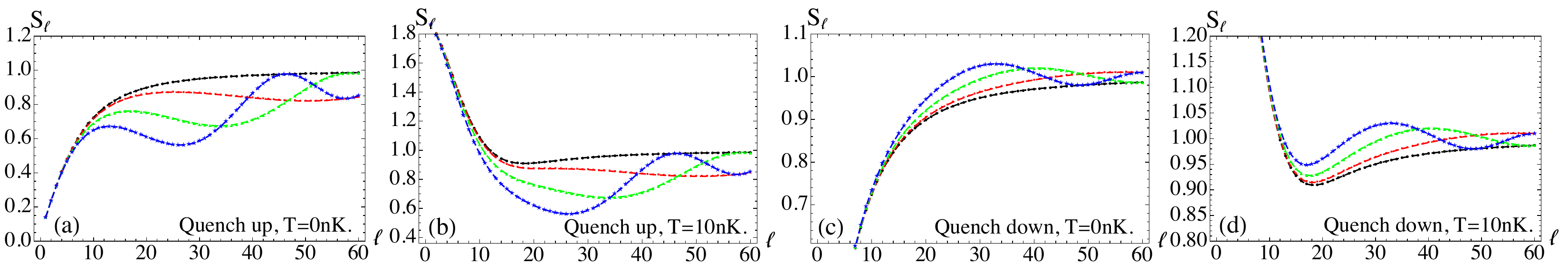}
    \caption{(Color online) BEC power spectrum $S_\ell(t)$, Eq.~(\ref{eq:BECpowerspectrum}), for various quenches, showing how the dynamical particle production is reflected in this observable.
    The different colors of the plots denote different times after the quench stops: $t=0\,\mathrm{s}$ (black), $t=0.3\,\mathrm{s}$ (red), $t=0.9\,\mathrm{s}$ (green), and $t=1.2\,\mathrm{s}$ (blue).
    The panels show
    (a) Quench up, $T=0\,\mathrm{K}$.
    (b) Quench up, $T=10\,\mathrm{nK}$.
    (c) Quench down, $T=0\,\mathrm{K}$.
    (d) Quench down, $T=10\,\mathrm{nK}$,
    with the other system parameters consistent with those of the Jia et al experiments~\cite{Jia2022} who achieved a shell-shaped BEC using a dual-species mixture (see Appendix~\ref{appendix:realisticparameter}). 
    In each case, $S_\ell$ shows oscillatory time dependence after the quench, although the character of such oscillations is different in the two cases, with the quench-up (quench-down) case showing oscillations below (above) the $t=0$ curve.  Note the overall amplitude of the oscillations is smaller in the quench-down case due to $|\beta_\ell|$ being smaller, as also seen in Fig.~\ref{fig:setup}.
    }
    \label{fig:SakharovOscillation}
\end{figure*}
\begin{equation}
    S_\ell(t)
    =
    \frac{\rho_0\varepsilon_\ell}{E_\ell^f}\left(
    1+N_\ell-\Delta N_\ell(t)
    \right)[1+2\nb(E_\ell^i)],
    \label{eq:BECpowerspectrum}
\end{equation}
where \(N_\ell=2|\beta_\ell|^2\) and \(\Delta N_\ell=2\Re[\alpha_\ell^*\beta_\ell e^{2iE_\ell^f t/\hbar}]\). 

The post-quench structure factor at different times is plotted in Fig.~\ref{fig:SakharovOscillation}. This is akin to the angular power spectrum in the analog CMB scenario. The power spectrum starts as a smooth curve in equilibrium, and evolves to an oscillatory one. As time progresses, the peaks move toward lower angular momentum. 

In the context of CMB physics, Sakharov oscillations refer to the acoustic multi-peak structure in the angular power spectrum. Our spherical-shell BEC model serves as a laboratory analog to this phenomenon: decomposing the shell density fluctuations into spherical harmonics maps the system onto the cosmological last scattering surface. In true cosmology, the acoustic peaks shift and compress over time due to cosmic expansion and the increasing distance to the surface of last scattering~\cite{Zibin2007}. Our shell BEC model captures similar effect of compressing peaks, but essentially abstracts away most of the cosmological details and only focus on the core analogy between phonon production and the cosmological power spectrum.

\begin{figure}[htbp]
    \centering
    \includegraphics[width=0.6\linewidth]{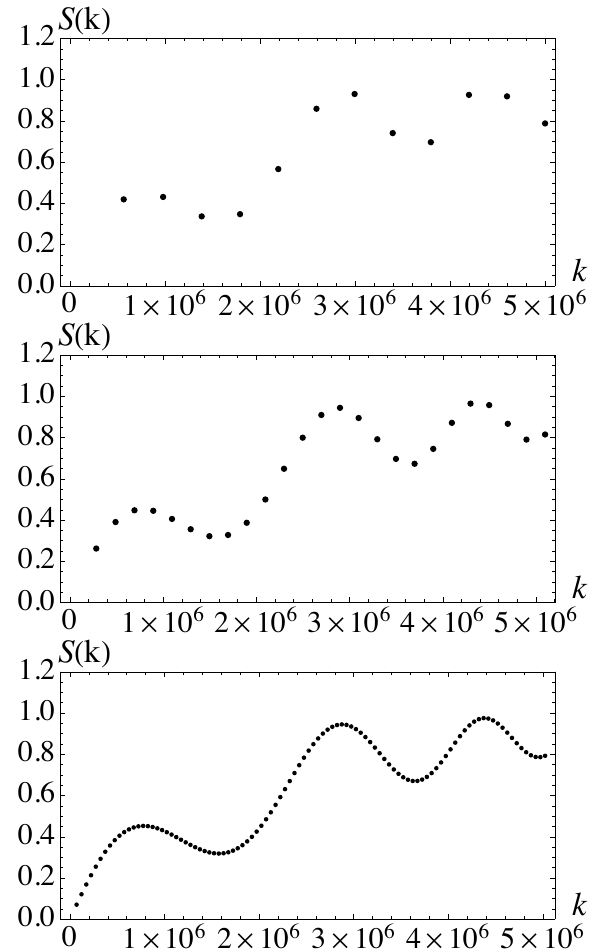}
    \caption{The three plots show the structure factor versus effective wave number 
    $k\equiv \sqrt{\ell(\ell+1)}/R$ (instead of $\ell$) at time $t=2\,\mathrm{s}$ with radii 
     $R = 0.18\,\mu\mathrm{m}$ (top panel), $0.28\,\mu\mathrm{m}$ (middle panel), and $0.58\,\mu\mathrm{m}$ (bottom panel). We see that these curves have the same shape, with the only difference being the point spacing (due to discreteness in $\ell$), showing that the structure factor is independent of $R$ when plotted vs. $k$.}
    \label{fig:curvaturedependence}
\end{figure}

One final question of interest concerns how our results for the particle production and structure factor of a quenched bubble BEC depend on the radius $R$. This question is particularly relevant because the radius of the bubble sets the curvature of the effective spacetime. With details relegated to Appendix~\ref{appendix:realisticparameter},we find that \(|\beta_\ell|^2\) and \(S_\ell\) do not depend explicitly on the radius $R$ (curvature), but instead on the spherical wave vector \(k=\sqrt{\ell(\ell+1)}/R\), which characterizes the physical wavelength of an excitation rather than its angular momentum quantum number \(\ell\). This result suggests that particle production preferentially occurs at particular physical wavelengths, rather than at particular values of \(\ell\). As shown in Fig.~\ref{fig:curvaturedependence}, results obtained for different radii collapse onto the same behavior when expressed as a function of \(k\). The role of curvature is therefore primarily to determine the discrete spectrum of allowed modes on the spherical surface, rather than to modify the underlying wavelength-dependent excitation dynamics.
  
This tells us that particle production vs. 
wavevector $k$ should be approximately universal, 
i.e., independent of $R$, as long as we work at fixed $\rho$.

\section{Entanglement}
\label{sec:entanglement}
In standard cosmology, primordial background fluctuations in the power spectrum are understood to originate from a two-mode squeezing process at the beginning of the universe~\cite{GlenzParker2009,Martin2016,Martin2017,Marchetal2020}. This is analogous to the creation of phonon field fluctuations via the modulation of a BEC system. Consequently, investigating the entanglement of quantum correlations in cold-atom platforms may provide valuable insights into the underlying entanglement structure of the cosmic power spectrum.

We note that prior theoretical work has explored the role of quantum correlations and entanglement in the context of analog gravity~\cite{Fuentes2010,Busch2014a,Busch2014b}. On the experimental side, such entanglement has been explicitly verified in BEC systems via the violation of the Cauchy-Schwarz inequality~\cite{Kheruntsyan2012,Wasak2014}.

To quantify the entanglement of particles produced during the interaction quench of a bubble BEC, we use logarithmic negativity (LN)~\cite{Serafini,Wang2020,WangWilde2020Negativity}, which is derived from taking the logarithm of the trace norm of the partial transpose of a density matrix. LN is an entanglement monotone linked to the Peres-Horodecki criterion~\cite{Peres:1996dw,Plenio:2005cwa,Simon:1999lfr} that remains computable for both pure and mixed states. Here, we specifically apply this measure to evaluate the bipartite quantum entanglement between particle pairs created during the quench. Specifically, we consider excitations in the conjugate modes $(\ell,m)$ and $(\ell,-m)$ defined by the spherical harmonic decomposition in Eq.~(\ref{modedecompositionrho}).

In the following, we will derive the partial transpose of the density matrix of our system. The first step is to define the input state, the initial equilibrium
bubble BEC, in terms of its mode operators. Let $\hat{\mathbf{B}}_{\ell m}^{(\text{in})}$ be the column vector comprising the \lq\lq in" creation and annihilation operators for a pair of spherical harmonic modes with angular indices $(\ell, m)$ and $(\ell, -m)$:
\begin{equation}
    \hat{\mathbf{B}}_{\ell m}^{(\text{in})} \equiv \left( \hat{b}_{\ell m}^{(\text{in})}, \hat{b}_{\ell m}^{(\text{in})\dagger}, \hat{b}_{\ell, -m}^{(\text{in})}, \hat{b}_{\ell, -m}^{(\text{in})\dagger} \right)^\top,
\end{equation}
with the \lq\lq out" mode vector
$\hat{\mathbf{B}}_{\ell m}^{(\text{out})}$ defined analogously.
The point here is that the quantum dynamics of the quench are encoded
in the  Bogoliubov transformation relating these states, as described by Eq.~(\ref{BogoliubovTransformation}).
We can therefore equivalently encode
these dynamics in the scattering matrix 
$\mathbf{S}_{(\mathbf{B},\ell,m)}$
that connects the initial and final mode operator vectors via:
$\hat{\mathbf{B}}_{\ell m}^{(\text{out})} = \mathbf{S}_{(\mathbf{B},\ell,m)} \cdot \hat{\mathbf{B}}_{\ell m}^{(\text{in})}$.
Our explicit result for the mode 
scattering matrix is:
\begin{equation}
    \mathbf{S}_{(\mathbf{B},\ell,m)} = 
    \begin{bmatrix}
        \alpha_\ell & 0 & 0 & (-1)^m\beta_\ell^* \\
        0 & \alpha_\ell^* & (-1)^m\beta_\ell & 0 \\
        0 & (-1)^{-m}\beta_\ell^* & \alpha_\ell & 0 \\
        (-1)^{-m}\beta_\ell & 0 & 0 & \alpha_\ell^*
    \end{bmatrix},
\end{equation}
where $\alpha_\ell$ and $\beta_\ell$ are the Bogoliubov coefficients that we have determined above.

Following Serafini, to obtain the covariance matrix we must connect the mode operators to the canonical position and momentum basis:
\begin{equation}
    \hat{x}_{\ell m} = \frac{1}{\sqrt{2}}(\hat{b}_{\ell m}+\hat{b}_{\ell m}^{\dagger}), \quad \hat{p}_{\ell m} = -\frac{i}{\sqrt{2}}(\hat{b}_{\ell m}-\hat{b}_{\ell m}^{\dagger}).
\end{equation}
The scattering matrix in this canonical basis, denoted as $\mathbf{S}_\ell$, is obtained via the transformation $\mathbf{S}_\ell = \mathbf{M}^{-1} \mathbf{S}_{(\mathbf{B},\ell,m)} \mathbf{M}^*$, where $\mathbf{M}$ is the appropriate basis-change matrix. 

We assume the system remains in a Gaussian state throughout the evolution, allowing it to be fully characterized by its first and second moments:
\begin{eqnarray}
    \mu_i &=& \langle \hat{r}_i \rangle, \\
    \sigma_{ij} &=& \langle \{ (\hat{r}_i - \mu_i), (\hat{r}_j - \mu_j) \} \rangle,
\end{eqnarray}
where $\hat{r} = (\hat{x}_{\ell m}, \hat{p}_{\ell m}, \hat{x}_{\ell, -m}, \hat{p}_{\ell, -m})^\top$ represent the canonical operators. 
In the present case, $\mu_i = 0$.  
The second moment, $\sigma$, defines the covariance matrix. For an initial thermal state (i.e., a finite-temperature BEC), the ``in" covariance matrix is proportional to the 
$4\times 4$ 
identity
matrix $\mathbb{I}_4$:
\begin{equation}
\boldsymbol{\sigma}_{\ell}^{(\text{in})} = (1 + 2n_B(E_\ell^i)) \mathbb{I}_4,
\end{equation}
where $n_B$ is the Bose-Einstein distribution function.   Since the quantum evolution during the quench is encapsulated by
the scattering matrix (and the parameters $\alpha_\ell$ and
$\beta_\ell$), the covariance matrix of the final system is simply
given by:
\begin{widetext}
\begin{eqnarray}
\label{eq:ourcov}
&&\boldsymbol{\sigma}_{\ell}^{(\mathrm{out})} = \mathbf{S}_\ell \cdot \boldsymbol{\sigma}_{\ell}^{(\mathrm{in})} \cdot \mathbf{S}_\ell^{\top} \\
&&= (1 + 2n_{\mathrm{B}}(E_\ell^i)) 
\begin{bmatrix}
|\alpha_\ell|^2 + |\beta_\ell|^2 & 0 &(-1)^m( \alpha_\ell^* \beta_\ell + \alpha_\ell \beta_\ell^* )& i(-1)^m(\alpha_\ell^* \beta_\ell - \alpha_\ell \beta_\ell^*) \\
0 & |\alpha_\ell|^2 + |\beta_\ell|^2 & i(-1)^m(\alpha_\ell^* \beta_\ell - \alpha_\ell \beta_\ell^*) & -(-1)^m(\alpha_\ell^* \beta_\ell + \alpha_\ell \beta_\ell^*) \\
(-1)^m(\alpha_\ell^* \beta_\ell + \alpha_\ell \beta_\ell^*) & i(-1)^m(\alpha_\ell^* \beta_\ell - \alpha_\ell \beta_\ell^*) & |\alpha_\ell|^2 + |\beta_\ell|^2 & 0 \\
i(-1)^m(\alpha_\ell^* \beta_\ell - \alpha_\ell \beta_\ell^*) & -(-1)^m(\alpha_\ell^* \beta_\ell + \alpha_\ell \beta_\ell^*) & 0 & |\alpha_\ell|^2 + |\beta_\ell|^2
\end{bmatrix}.
\nonumber 
\end{eqnarray}
\end{widetext}
The next step is to connect the covariance matrix to the LN  measure of  entanglement
$E_{\mathcal{N}}$ between the modes $(\ell, m)$ and $(\ell, -m)$.  The idea of LN is 
that the uncertainty principle, required for all quantum systems, relies on the condition
$\boldsymbol{\sigma} + i\Omega\geq 0$, with 
$\Omega$ the symplectic form.  For a two-mode system, this condition is equivalent to stating that the symplectic eigenvalues $\nu_i$
of $\boldsymbol{\sigma}$ (obtained from the eigenvalues of $i\Omega \boldsymbol{\sigma}$, which are $\pm \nu_i$) must satisfy $\nu_i\geq 1$.

To get $E_{\mathcal{N}}$, we consider the symplectic eigenvalues of the partial-transposed covariance matrix $\boldsymbol{\sigma}^{\rm PT} = T \boldsymbol{\sigma} T$ with $T = {\rm diag}(1,1,1,-1)$ which implements a partial transpose operation (mapping $p\to -p$ for one mode).  The 
idea of the partial transpose operation is as follows: If the original system is separable, then the partial transpose operation constructs the covariance matrix of a different quantum system still obeying $\tilde{\nu}_i\geq 1$
(with the tilde indicating the partial transpose operation).  If the original system is entangled, however, then the covariance matrix of
the partial transpose system is unphysical, with one of the two symplectic eigenvalues becoming less than unity. 

The LN entanglement measure of a 
two-mode system depends only  on the lower symplectic eigenvalue $\tilde{\nu}_-$:
\begin{equation}
    E_{\mathcal{N}}[\ell] = \max \Big[ 0, -\log_{2} \tilde{\nu}_- \Big],
\end{equation}
which increases monotonically as 
$\tilde{\nu}_-$ reduces below unity. For our system with
covariance matrix
Eq.~(\ref{eq:ourcov}), this reduces to:
\begin{eqnarray}
&&    E_{\mathcal{N}}[\ell]  \label{LogNegFormula}
    \\
&&    = \max \Big[ 0, -\log_{2} \left( (1+2n_{\text{B}}(E_\ell^i)) \left( \sqrt{1+|\beta_{\ell}|^2} - |\beta_{\ell}| \right)^{2} \right) \Big],
    \nonumber
\end{eqnarray}
depending on the system temperature
and the particle production number $|\beta_{\ell}|^2$.  A nonzero $E_{\mathcal{N}}[\ell]$ implies that
modes $(\ell,m)$ are entangled, essentially in a two-mode squeezed state.  

\begin{figure}[!h]
  \centering
    \includegraphics[width=0.8\linewidth]{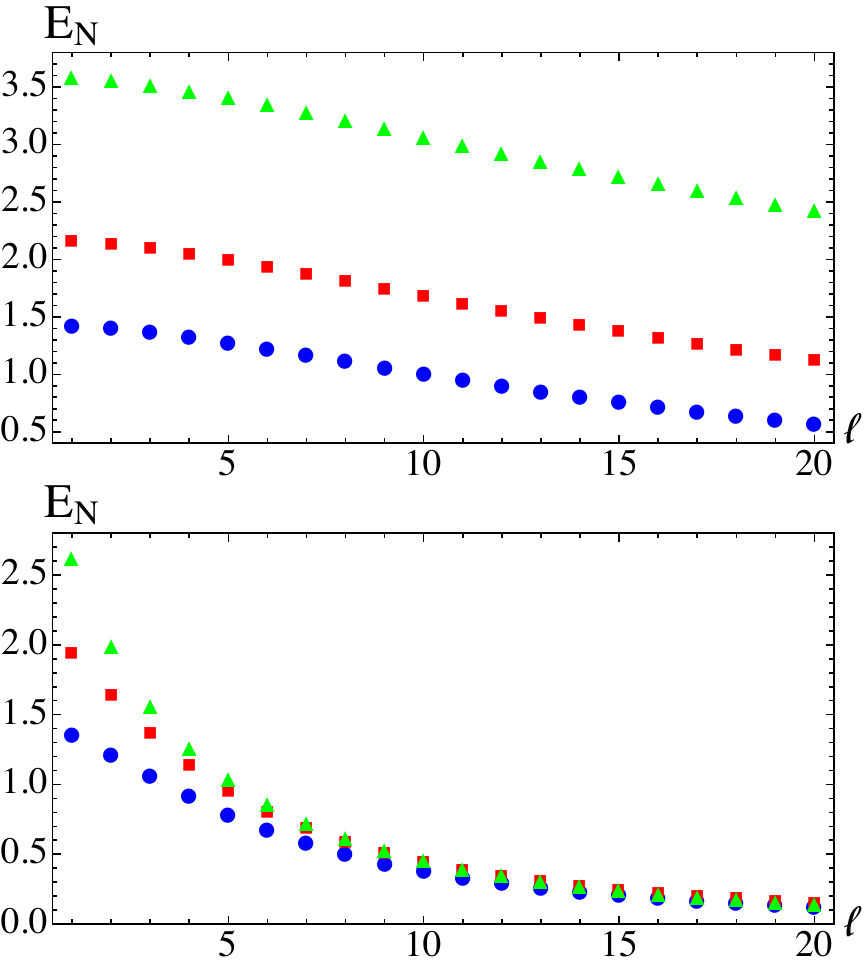}
  \caption{(Color online)
  Logarithmic Negativity measure of the entanglement between particles with quantum number $\ell$ and $(m,-m)$ for the vacuum initial state ($n_B=0$). The colors represent different quench durations \(t_f=20\,\mathrm{ms}\) (blue circles), \(t_f=30\,\mathrm{ms}\) (red
  squares), and \(t_f=50\,\mathrm{ms}\) (green triangles). We assume $T=0\,\mathrm{K}$ for both panels. Thus, in the low-$T$ limit, the logarithmic negativity follows qualitatively the same trend as the particle production $|\beta_\ell|^2$.}
  \label{fig:LogNegNoTemp}
\end{figure}

\begin{figure}[!h]
  \centering
  \includegraphics[width=0.8\linewidth]{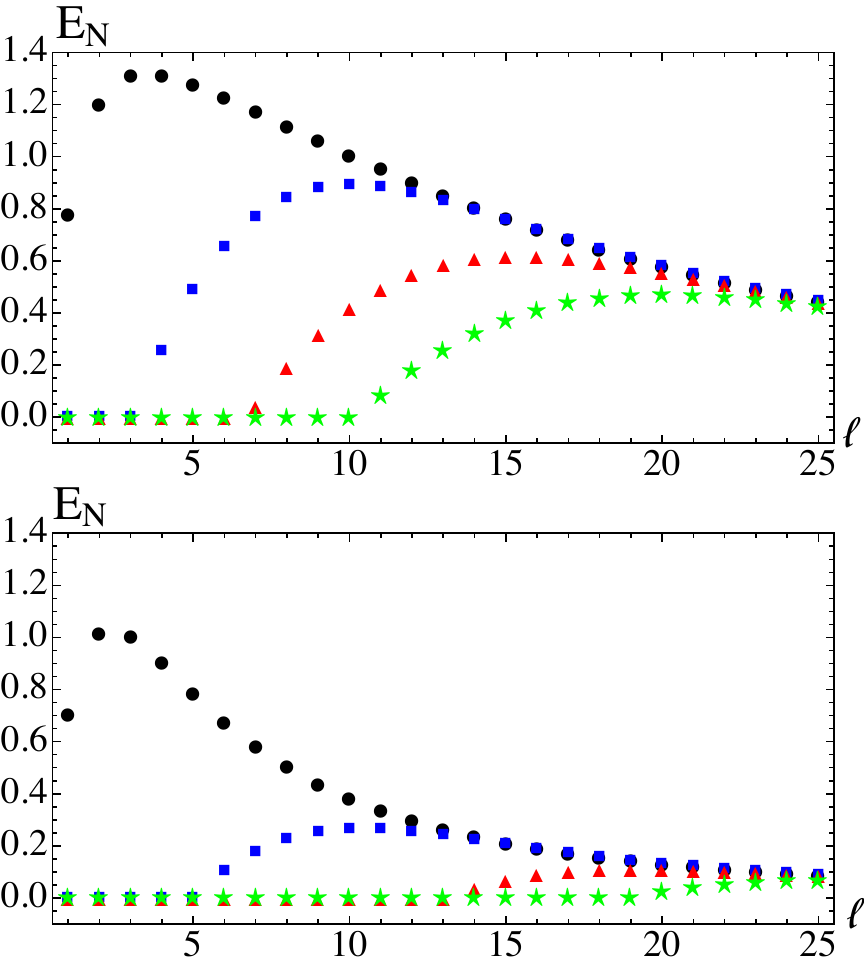}
  \caption{(Color online) Logarithmic Negativity at nonzero  temperature, \(T=1\,\mathrm{nK}\) (black circles), \(T=5\,\mathrm{nK}\) (blue squares), \(T=10\,\mathrm{nK}\) (red triangles), and \(T=15\,\mathrm{nK}\) (green stars).
  The quench duration is  \(t_f=20\,\mathrm{ms}\) for all curves.  We see that as the temperature increases, the entanglement of lower-$\ell$ modes is strongly suppressed.}
  \label{fig:LogNegTemp10}
\end{figure}

The resulting entanglement as a 
function of $\ell$ for a quenched spherical-shell BEC is illustrated in Figs.~(\ref{fig:LogNegNoTemp}) and (\ref{fig:LogNegTemp10}). We compare cases with and without ambient temperature, mirroring recent investigations into entanglement negativity within Hawking radiation~\cite{Agullo:2021vwj,Brady:2022ffk} and analog inflationary models~\cite{Bhardwaj2024EntanglementBEC}.

\subsection{Measuring Entanglement}

It is of considerable interest to find 
direct experimental probes of entanglement.
One way to do this, discussed in Ref.~\cite{Bhardwaj2024EntanglementBEC}, is
to use the fact that the covariance matrix
is determined, in a Gaussian system, by
various correlation functions (e.g., 
density-density, phase-phase and density-phase
correlators).  Therefore, an experiment that measures these can reconstruct the covariance matrix and,
hence, the log neg.  This method is limited by 
the difficulty of measuring these correlators.  

In this section, we study another method that
can extract the particle production number, 
$|\beta_\ell|^2$ and, hence, probe the system entanglement.
Our proposed method is inspired by the earlier work
of Chen et al, who experimentally studied 
quasiparticle production under a modulation instability in a 2D BEC~\cite{Chen2021Quasiparticle}.  Their measurements are based on the density noise power spectrum \(S_k=\braket{|\delta k|^2}/N\). They defined the rescaled structure factor as:

\begin{equation}
    \tilde{S}_\ell(t) =
    \frac{S_\ell(t)}{\rho_0\varepsilon_\ell / E_\ell^f} = \left( 1 + N_\ell - \Delta N_\ell(t) \right).
    \label{SkTilde}
\end{equation}

The time-dependent $\Delta N_\ell$ is oscillatory, and can be written as:
\be
\Delta N_\ell(t) =2 |\alpha_{\ell}||\beta_{\ell}|
\cos\Big( \frac{2E_\ell^f t}{\hbar} +\phi_\ell
\Big),
\label{eq:deltaNformula} 
\ee
with $\phi_\ell$ a phase that depends on $\alpha_{\ell}$ and $\beta_{\ell}$. 

The period of the oscillatory behavior enables them to get the phonon energy $E_\ell^f$ for calculating the rescaled structure factor. The amplitude could be used to obtain the particle production density $N_\ell$ and the pair correlation $\Delta N_\ell$:
\begin{subequations}
    \begin{equation}
     N_\ell = 2|\beta_\ell|^2 =\frac{\tilde{S}_{\text{max}} + \tilde{S}_{\text{min}}}{2} - 1,
        \label{NBetaSquare}
    \end{equation}
    \begin{equation}
          |\Delta N_\ell| =2 |\alpha_{\ell}||\beta_{\ell}| = \frac{\tilde{S}_{\text{max}} - \tilde{S}_{\text{min}}}{2}.
    \end{equation}
\end{subequations}
By the Peres-Horodecki principle, the correlation is non-classical if and only if $|\Delta N_\ell|>N_\ell$. 

\begin{figure}
    \centering
    \includegraphics[width=\linewidth]{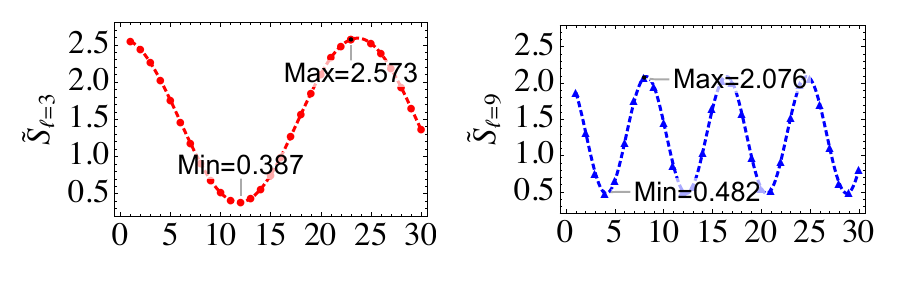}
    \caption{Examples of \(\tilde{S}_\ell(t)\), Eq.~(\ref{SkTilde}), for \(\ell=3,9\).
    These are sinusoidal waves, with maximum and minimum. The particle production amplitude can be obtained using Eq.~(\ref{NBetaSquare}): \(|\beta_3|^2=0.24\), and \(|\beta_9|^2=0.14\), which are in agreement with the top panel of Fig.~\ref{Betasquare}, with duration $t_f=20\,\mathrm{ms}$.
    }
    \label{fig:SkTildeTime}
\end{figure}

Typical examples of $\tilde{S}_\ell(t)$ are shown in Fig.~\ref{fig:SkTildeTime}. 
The data shown here is identical to that in Fig.~\ref{fig:SakharovOscillation}, but plotted versus time. Neglecting thermal noise from the Bose factor, the rescaled structure factor gives $|\beta_\ell|^2$, which is required to calculate the logarithmic negativity in Eq.~(\ref{LogNegFormula}).
In our calculation, however, the structure factor includes a thermal correction term, $(1+2n_B)$, which becomes non-negligible at long wavelengths and higher temperatures.
This thermal effect degrades entanglement, as illustrated in Fig.~\ref{fig:LogNegTemp10}. While the entanglement criterion used by Chen's group---$(\tilde{S}_{\text{max}} - \tilde{S}_{\text{min}})/2>(\tilde{S}_{\text{max}} + \tilde{S}_{\text{min}})/2 - 1$---remains valid even with this hidden $(1+2n_B)$ factor, it could overestimate the quantum correlations. In the long-wavelength, finite-temperature regime, the actual non-classical correlations are scarcer than those observed by experiment. To address this, we propose a different way to compute the $|\beta_\ell|^2$ using the original structure factor, with thermal contribution factored out:
\begin{equation}
    \frac{S_{\text{max}}-S_{\text{min}}}{S_{\text{max}}+S_{\text{min}}}=\frac{2\sqrt{1+|\beta_\ell|^2}|\beta_\ell|}{1+2|\beta_\ell|^2}.
    \label{eq:beta_from_structure_factor}
\end{equation}
The right side is a monotonic function of $|\beta_\ell|$, so measurement of the left side uniquely determines a particle production $|\beta_\ell|^2$, which is the true particle production density $N_\ell$ without the thermal distribution.

\section{Concluding Remarks}
\label{sec:concl}
In this paper, we 
proposed spherical-shell (or bubble) BECs as a platform for analog gravity experiments.  We showed how an interacting spherical-shell BEC can
realize an intrinsically curved analog
spacetime and that a sudden increase
or decrease of the interaction 
strength can realize an analog 
\lq\lq big crunch\rq\rq or \lq\lq big bang\rq\rq.

Our calculations relied on the
development of the Bogoliubov theory for a spherical-shell BEC.  Starting from a
model of a 3D trapped spherical-shell 
BEC, we  carefully derived an effective
2D theory in the thin-shell limit (see Appendix~\ref{appendix:dimensional_reduction}).  
Notably,
our general analysis included the possibility
of other types of quench, including
changes in the sphere radius $R(t)$
and spherical-shell thickness $\Delta R(t)$, although our main focus was 
on interaction quenches.

Using  the Madelung representation, we derived a phonon mode equation suitable for studying non-equilibrium dynamics
of a quenched spherical-shell BEC.  This dynamics
is reflected in quantum particle production after such interaction quenches (which correspond, within the cosmology analog, to quenches in the scale factor).  We showed how this particle
production, characterized by the quantity $|\beta_\ell|^2$ (with $\ell$ the spherical quantum number), leads to a time-dependent term in the boson density-density correlation function or structure factor, providing a specific prediction for our scenario.  Notably, the time evolution of the structure factor in the post-quench regime exhibits a multi-peak structure that resembles cosmic microwave background (CMB) Sakharov oscillations.

Indeed, as we also discussed, 
the abovementioned quantum particle
production could be thought of as
a two-mode squeezing process that
entangles states in spherical-harmonic spatial wavefunctions
with quantum numbers 
$(\ell,m)$ and $(\ell,-m)$.  We 
showed how the amount of entanglement 
could be quantified using the logarithmic negativity (log neg) measure.  Since the log neg and structure factor both depend on $|\beta_\ell|^2$, a measurement of
the structure factor therefore also provides information on entanglement, 
as also found in the experiments of 
Ref.~\cite{Chen2021Quasiparticle}.

We note here some connections to 
related work. 
Investigating a static spherical-shell BEC with a time-dependent interaction strength makes this work overlap with Refs.~\cite{TolosaSimeon2022,SanchezKuntz2022}.
There they considered a time-dependent s-wave scattering length in a quasi-2D pancake-type geometry whose trapping potential was chosen such that, after an appropriate coordinate transformation, they could realize a 2-sphere geometry (as well as flat and hyperbolic geometries).
This system was experimentally realized in Ref.~\cite{Viermann2022} and used to simulate quantum fluctuations in analog expanding curved spacetimes; building directly on that setup, two of the current authors and collaborators studied the field propagation in an inhomogeneous disk BEC~\cite{Chunnetal2026}.
Inflationary physics has also been studied by one of the current authors and collaborators in expanding ring-shaped BECs, while keeping the interaction strength fixed~\cite{Bhardwaj2021Inflationary,Bhardwaj2024EntanglementBEC}.

The Bogoliubov theory discussed here for fluctuations about a spatially isotropic background condensate is closely related to what was developed in \cite{Tononi2019} and the review \cite{Tononi2024}.
A key distinction in our exposition is that we work in the canonical operator formalism rather than the path integral formalism. Furthermore, we consider time-dependent couplings in the Hamiltonian and investigate logarithmic negativity as an entanglement measure in this setting, including potential methods for its measurement.

Our work opens up several promising directions for future study. The dynamical equations we derived can
be used to study other types of quench (such as a time-dependent radius $R(t)$ or thickness $\Delta R(t)$, as mentioned above), but also different
time-dependences beyond the exponential quenches 
studied here.  For example, one could study a 
periodic quench, which transforms our mode equation into a Mathieu-type equation. Periodic quenching of a BEC cloud is associated with pattern formation and granulation~\cite{Zhang2020,Kwon2021,Liebster2025}.

Our model idealizes the bosonic cloud to be a uniform and thin spherical shell. In experiments like the Cold Atom Lab (CAL), the atomic cloud forms an ellipsoid with an inhomogeneous background density rather than an ideal sphere. Thus, the finite-thickness effects and potential radial excitations could alter the dynamics. Developing a generalized Bogoliubov framework that accounts for these geometric and density variations will greatly refine future theoretical predictions. Furthermore, incorporating rotation into this shell framework would allow us to study vortex dynamics and topological defect structures on curved surfaces.

More broadly, these results establish spherical-shell BECs as a platform for studying quantum field dynamics in curved, time-dependent spacetimes. By connecting particle production and entanglement to measurable density correlations, this system provides a way to probe the quantum nature of analog cosmological phenomena. At the same time, ideas from quantum field theory and cosmology can offer new perspectives on cold-atom many-body physics, particularly on how entanglement can be detected experimentally. Analog-gravity systems therefore offer rich ground for multi-way cross-fertilization between sub-disciplines: employing cold atomic settings as a highly tunable platform for exploring cosmology, and infusing concepts from cosmology in probing phenomena manifest in quantum many-body systems.

\section{Acknowledgement}

We thank James A. Chunn and Chris M. Herdman for useful discussions.
R.Z. and D.E.S. acknowledge support from the National Science Foundation under Grants PHY-2208036 and PHY-2608230.
B.R. and S.V. acknowledge support by the National Aeronautics and Space Administration (NASA) Science Mission Directorate, Division of Biological and Physical Sciences (BPS) through ROSES-22 as well as multiple Jet Propulsion Laboratory (JPL) Research Support Agreements.
B.R. and N.G. gratefully acknowledge funding from the DLR Space Administration with funds provided by the Federal Ministry for Economic Affairs and Climate Action (BMWK) under grant numbers 50WM2451 (QUANTUMANIA) and from the Deutsche Forschungsgemeinschaft (DFG) through SFB 1227 (DQ-mat) within Project A05, Germany’s Excellence Strategy (EXC-2123 QuantumFrontiers Grants No. 390837967).
Z.M. acknowledges support from the Australian Government through Australian Research Council (ARC) Project No. DP260102306. 

\bibliography{references}

\appendix

\section{Dimensional Reduction}
\label{appendix:dimensional_reduction}

We are only interested in trapping potentials, $V(\mathbf{r},t)$, in which the bosons are tightly confined to a thin spherically-symmetric shell of radius $R(t)$ and thickness $\Delta R(t)$ with $R(t)/\Delta R(t) \gg 1$.
We assume in all cases that the temperature is sufficiently low and the quench protocol is sufficiently slow as to not populate higher energy radial excitations.
Treating the radial dynamics as frozen-out, we work with an effective theory in terms of field operators defined on the surface of the sphere. 
In the Schr\"odinger picture we define
\begin{align}
    \hat\Psi_S(\theta,\varphi,t)
    &\equiv
    \int_0^\infty dr \, r f^*(r,t)
    \hat\Psi(r,\theta,\varphi)
    ,
    \label{eq:2_sphere_field_operators}
\end{align}
where the radial distribution $f(r,t)$ is strongly-peaked about the shell radius, has a characteristic width on the order of the shell thickness, and is normalized such that $1=\int_0^\infty dr |f(r,t)|^2$. These field operators are dimensionless and obey equal-time bosonic commutation relations on the surface of the sphere, Eq.~(\ref{eq:2_sphere_bosonic_commutation_relations}).
In the thin-shell limit, $R(t)/\Delta R(t) \gg 1$, we assume that Eq.~(\ref{eq:2_sphere_field_operators}) can approximately be inverted
\begin{align}
    \hat\Psi(r,\theta,\varphi) &\approx \frac{f(r,t)}{r}
    \hat\Psi_S(\theta,\varphi,t)
    .
    \label{eq:dimensionally_reduced_field_operators}
\end{align}

For concreteness, let us consider a Gaussian radial distribution \cite{Tononi2025}:
\begin{align}
    f(r,t) = \frac{1}{\pi^{1/4}\sqrt{\Delta R(t)}} \exp\left[-\frac{1}{2} \left(\frac{r-R(t)}{\Delta R(t)}\right)^2\right]
    ,
    \label{eq:radial_distribution}
\end{align}
which is properly normalized if we work asymptotically as $R(t)/\Delta R(t) \to \infty$.
This radial distribution arises naturally in the thin-shell limit of the spherically symmetric bubble trap, Eq.~(\ref{eq:V_bubble}).
For sufficiently large detunings, $\Delta(t)$, the bubble trap potential is well approximated by a radially shifted harmonic oscillator with frequency $\omega_\mathrm{sh}(t) \equiv \omega_0 \sqrt{\Delta(t)/\Omega}$ and radius $R(t) = s_l \sqrt{\Delta(t)}$.
Assuming the temperature of the system is far below the gap to higher energy radial excitations, $k_{\mathrm{B}} T \ll \hbar \omega_\mathrm{sh}(t)$, and the rate at which the parameters in the Hamiltonian vary is much slower than $\omega_\mathrm{sh}(t)$, we can take $f(r,t)/r$ to be the radial component of the oscillator's ground state wavefunction.
In this case $f(r,t)$ is given by Eq.~(\ref{eq:radial_distribution}) with $\Delta R(t) = \sqrt{\hbar / M \omega_\mathrm{sh}(t)}$~\cite{Tononi2025}.

We note that the Schr\"odinger field operators on the surface of the sphere are static in the thin-shell limit, and hence one can simply replace $\hat\Psi_S(\theta,\varphi,t) \to \hat\Psi(\theta,\varphi)$. To see this, differentiate Eq.~(\ref{eq:2_sphere_field_operators}) and then insert Eq.~(\ref{eq:dimensionally_reduced_field_operators}) in the thin-shell limit to obtain $i\hbar\partial_t \hat\Psi_S(\theta,\varphi,t) \approx - \gamma(t) \hat\Psi_S(\theta,\varphi,t)$ with $\gamma(t) \equiv \int_0^\infty dr \left[ i\hbar\partial_t f(r,t) \right]^* f(r,t)$. The result follows by showing that $\gamma(t)=0$, which can be established from the real-valuedness and normalization of the distribution.

We can now perform the dimensional reduction. Starting with the Hamiltonian in Eq.~(\ref{eq:H_3D}), we insert Eq.~(\ref{eq:dimensionally_reduced_field_operators}) in the thin-shell limit which yields Eq.~(\ref{eq:H_2D_eff}) where the couplings are given by~\cite{Tononi2025}
\begin{align}
    \mu_\mathrm{2D}(t) &\sim \mu - \frac{\hbar^2}{2M\Delta R(t)^2}
    ,\\
    \lambda(t) &\sim 
    \frac{1}{\sqrt{2\pi}}
    \frac{U_0(t)}{R(t)^2 \Delta R(t)}
    ,
\end{align}
as $R(t)/\Delta R(t) \to \infty$.

\section{Bogoliubov Theory of Spherical-Shell BECs}
\label{appendix:Bogo_2-sphere}

In this appendix, we present an alternative and more detailed route to reach the linearized Heisenberg equations of motion and mode expansions for the density and phase fluctuation operators, Eq.~(\ref{eq:rho_phi_fluc_EOM}) and Eq.~(\ref{modeexpansion}) in the main text respectively.

\subsection{Linearized Equations of Motion for Fluctuations}
Using Eq.~(\ref{eq:H_2D_eff}) to generate dynamics on the surface of the sphere, we define Heisenberg operators as
\begin{align}
    \hat\Psi(\theta,\varphi,t) \equiv \hat U^\dag(t) \hat\Psi(\theta,\varphi) \hat U(t)
    \label{eq:Heisenberg},
\end{align}
where
\begin{align}
    \hat U(t) = \mathcal{T} \exp \left( - \frac{i}{\hbar} \int_{0}^{t} dt' \hat H_\text{2D}(t') \right)
    \label{eq:U},
\end{align}
is the unitary time evolution operator with $\mathcal{T}$ representing time-ordering. The Heisenberg equation of motion (EOM) for the dimensionless field operator on the surface of the sphere is given by an operator version of the Gross-Pitaevskii (GP) equation:
\begin{align}
    i\hbar \partial_t \hat\Psi
    &=
    \left( \frac{L^2}{2MR(t)^2} - \mu_\mathrm{2D}(t) + \lambda(t) \hat\Psi^\dag \hat\Psi \right) \hat\Psi
    .
\end{align}

Assuming $\mathrm{U}(1)$ spontaneous symmetry breaking, and considering temperatures far below the transition, we expand the field operator in fluctuations about the background condensate,
\begin{align}
    \hat\Psi(\theta,\varphi,t)
    =
    \Psi_0(\theta,\varphi,t) + \hat \psi(\theta,\varphi,t)
    ,
\end{align}
where the mean field, $\Psi_0 \equiv \langle \hat\Psi \rangle$, obeys the GP equation neglecting $\mathcal{O}(\hat\psi^2)$ terms. Up to linear terms, the Heisenberg EOM for the field fluctuations is given by
\begin{align}
    i\hbar \partial_t \hat \psi
    & =
    \left( \frac{L^2}{2MR(t)^2} - \mu_\mathrm{2D}(t) + 2 \lambda(t) |\Psi_0|^2 \right) \hat \psi
    \nonumber\\
    &+ \lambda(t) \Psi_0^2 \hat \psi^\dag
    + \mathcal{O}(\hat \psi^2)
    .
    \label{eq:EOM_fluctuations}
\end{align}

In this work, we consider spatially isotropic condensate wavefunctions, $\Psi_0(\theta,\varphi,t)=\Psi_0(t)$, where the GP equation is readily solved. Using the Madelung representation, the condensate wavefunction is $\Psi_0(t) = \exp(i \phi_0(t)) \sqrt{\rho_0}$ where the angular condensate density, $\rho_0$, is constant and the phase of the condensate is given by
\begin{align}
    \phi_0(t) =
    - \frac{1}{\hbar} \int_{0}^{t} dt' \left[ - \mu_\mathrm{2D}(t') + \lambda(t') \rho_0 \right]
    ,
\end{align}
where we set the initial phase to zero without loss of generality.

For a spatially isotropic condensate, it is useful to expand the field fluctuations in spherical harmonics:
\begin{align}
    \label{eq:mode_expansion_psi}
    \hat\psi(\theta,\varphi,t) = \sum_{\ell m} Y_{\ell m}(\theta,\varphi) \hat{a}_{\ell m}(t)
    ,
\end{align}
where $\sum_{\ell m} = \sum_{\ell=0}^{\infty} \sum_{m=-\ell}^{\ell}$.
After inserting the mode expansion into Eq.~(\ref{eq:EOM_fluctuations}) one finds
\begin{align}
    i\hbar \partial_t \hat{a}_{\ell m}(t)
    & =
    \xi_{\ell}(t) \hat{a}_{\ell m}(t)
    + (-1)^{m} \Delta_{\ell}(t) \hat{a}_{\ell,-m}^\dag(t)
    \nonumber\\
    &+ \mathcal{O}(\hat{a}_{\ell m}^2)
    ,
    \label{eq:EOM_fluctuations_Ylm}
\end{align}
where we define
\begin{subequations}
\label{eq:Bogo_energies}
\begin{align}
    \xi_{\ell}(t) &\equiv \varepsilon_\ell(t) - \mu_\mathrm{2D}(t) + 2 \lambda(t) \rho_0
    ,
    \\
    \Delta_{\ell}(t) &\equiv
    \lambda(t)
    \rho_0
    e^{i2\phi_0(t)}
    .
\end{align}
\end{subequations}
with
\begin{align}
    \varepsilon_\ell(t) \equiv \frac{\hbar^2 \ell(\ell+1)}{2MR(t)^2}
    ,
\end{align}
being an eigenvalue of the operator $L^2/2MR(t)^2$.
The formal solution to the linearized EOM, Eq.~(\ref{eq:EOM_fluctuations_Ylm}), is given by a Bogoliubov transformation:
\begin{align}
    \label{eq:time_evolved_mode_operators}
    \hat{a}_{\ell m}(t) = u_{\ell}(t) \hat{a}_{\ell m} + (-1)^{m} v^*_{\ell}(t) \hat{a}^\dag_{\ell,-m}
    ,
\end{align}
where $\hat{a}_{\ell m}\equiv\hat{a}_{\ell m}(0)$ are the initial mode fluctuation operators, and the coefficients $u_{\ell}(t)$ and $v_{\ell}(t)$ obey the Bogoliubov equations:
\begin{subequations}
\label{eq:Bogo}
\begin{align}
    i\hbar \partial_t u_{\ell}(t) & = \xi_{\ell}(t) u_{\ell}(t) + \Delta_{\ell}(t) v_{\ell}(t)
    ,
    \\
    i\hbar \partial_t v_{\ell}(t) & = - \xi_{\ell}(t) v_{\ell}(t) - \Delta^*_{\ell}(t) u_{\ell}(t)
    ,
\end{align}
\end{subequations}
with initial conditions $u_{\ell}(0)=1$ and $v_{\ell}(0)=0$.
It is straightforward to show that the coefficients obey the normalization condition $1=|u_{\ell}(t)|^2-|v_{\ell}(t)|^2$.

\subsection{Density and Phase Fluctuations}

Working in the quantum acoustic representation, density and phase fluctuation operators are defined as~\cite{Barcelo2011}:
\begin{align}
    \label{eq:rho_phi_fluc}
    \hat\rho
    &\equiv
    \Psi_0^* \hat \psi + \Psi_0 \hat \psi^\dag
    \quad,\quad
    \hat\phi
    \equiv
    - \frac{i}{2} \left(\frac{1}{\Psi_0} \hat \psi - \frac{1}{\Psi_0^*} \hat \psi^\dag \right)
    .
\end{align}
These operators are hermitian and obey equal-time canonical commutation relations on the surface of the sphere, Eq.~(\ref{eq:CCR_rho_phi}).
[Note: if one instead starts with the Madelung representation, Eq.~(\ref{Madelung}), these are the expressions that appear after a naive series expansion working to linear order in the fluctuations.]
Using the GP equation for the condensate and Eq.~(\ref{eq:EOM_fluctuations}) for the field fluctuations, one can compute the linearized Heisenberg EOM for the density and phase fluctuations, which gives Eq.~(\ref{eq:rho_phi_fluc_EOM}) in the main text.

We now insert the mode expansion, Eq.~(\ref{eq:mode_expansion_psi}), and the time-evolved mode operators, Eq.~(\ref{eq:time_evolved_mode_operators}), into Eq.~(\ref{eq:rho_phi_fluc}) to obtain
\begin{subequations}
\label{eq:density_phase_spherical_harmonic_expansion}
\begin{align}
    \hat\rho(\theta,\varphi,t)
    &=
    \sum_{\ell m} \Big[ Y_{\ell m}(\theta,\varphi) P_{\ell}(t) \hat{a}_{\ell m}
    + Y_{\ell m}^*(\theta,\varphi) P_{\ell}^*(t) \hat{a}_{\ell m}^\dag \Big]
    , \\
    \hat\phi(\theta,\varphi,t)
    &=
    \sum_{\ell m} \Big[ Y_{\ell m}(\theta,\varphi) Q_{\ell}(t) \hat{a}_{\ell m}
    + Y_{\ell m}^*(\theta,\varphi) Q_{\ell}^*(t) \hat{a}_{\ell m}^\dag \Big]
    ,
\end{align}
\end{subequations}
where the density and phase mode coefficients are given by linear combinations of the Bogoliubov coefficients,
\begin{subequations}
\begin{align}
    P_{\ell}(t) &\equiv \Psi_0^*(t) u_{\ell}(t) + \Psi_0(t) v_{\ell}(t)
    , \\
    Q_{\ell}(t) &\equiv - \frac{i}{2\Psi_0(t)} u_{\ell}(t) + \frac{i}{2 \Psi_0^*(t)} v_{\ell}(t)
    ,
\end{align}
\end{subequations}
and obey the EOM,
\begin{subequations}
\label{eq:mode_eqs}
\begin{align}
    \partial_t P_{\ell}(t) &= \left( \frac{\hbar \rho_0 \, \ell(\ell+1)}{MR(t)^2} \right) Q_{\ell}(t)
    ,
    \\
    -\frac{\hbar}{\lambda(t)} \partial_t Q_{\ell}(t)
    &= \left( 1 + \frac{\hbar^2 \, \ell(\ell+1)}{4 M \rho_0 R(t)^2 \lambda(t)} \right) P_{\ell}(t)
    ,
\end{align}
\end{subequations}
with initial conditions $P_{\ell}(0)=\sqrt{\rho_0}$ and $Q_{\ell}(0)=-\frac{i}{2\sqrt{\rho_0}}$.
It is straightforward to show that the coefficients obey the Wronskian condition: $i=P_{\ell}(t)Q_{\ell}^*(t) - P_{\ell}^*(t)Q_{\ell}(t)$.

\subsection{Mode Expansions for an Initial Thermal State}

In this work, we consider the initial system to be in thermal equilibrium at temperature $T$ and described by the density operator $\frac{1}{Z}e^{- \hat{H}_\text{2D}(0) / k_{\mathrm{B}}T}$,
where $Z=\mathrm{Tr}(e^{- \hat{H}_\text{2D}(0) / k_{\mathrm{B}}T})$ is the partition function.
For consistency with the rest of the section, the linearized Heisenberg EOM are generated by working with the Hamiltonian expanded to quadratic order in fluctuations about the condensate wavefunction.
Furthermore, since the initial density operator is thermal, we treat $\partial_t \Psi_0(0) = 0$. The time-independent GP equation leads to a relationship between the initial chemical potential and the angular condensate density, $\mu_\mathrm{2D}(0) = \lambda(0) \rho_0$, and results in an initial Hamiltonian operator of the form
\begin{align}
    \hat{H}_\text{2D} (0)
    &= -\frac{1}{2} \lambda_0 4\pi \rho_0^2
    + \sum_{\ell m}
    \Bigg[
    \left(\varepsilon_\ell^0 + \lambda_0 \rho_0\right) \hat{a}_{\ell m}^\dag \hat{a}_{\ell m}
    \nonumber\\
    &+ \frac{1}{2}
    \left(
    (-1)^{m}\lambda_0\rho_0
    \, 
    \hat{a}_{\ell m}^\dag \hat{a}_{\ell,-m}^\dag + \text{h.c.}
    \right)
    \Bigg]
    + \mathcal{O}(\hat{a}_{\ell m}^3)
    ,
\end{align}
where we have defined $\lambda_0 \equiv \lambda(0)$ and $\varepsilon_\ell^0 \equiv \varepsilon_\ell(0)$.

Diagonalizing the quadratic portion of the initial Hamiltonian is straightforward using Bogoliubov transformations~\cite{Blaizot1986}. The only subtlety comes from the $\ell,m=0,0$ mode, which becomes a squared quadrature operator (i.e. ``$\hat p^2$" or ``$\hat q^2$") as opposed to a quasiparticle number operator~\cite{Lewenstein1996,PethickSmith}.
Here we employ the standard Bogoliubov approximation by assuming a BEC of the $\ell,m=0,0$ mode and treat fluctuations in this mode as so small as to be neglected: $\hat{a}_{0,0} \to 0$.
This then gives a relation between the angular condensate density and the number of condensed atoms: $N_{0,0} = 4\pi \rho_0$.
After making the Bogoliubov approximation, neglecting cubic and quartic fluctuation terms, and performing a Bogoliubov transformation on the non-condensed modes, one obtains the effective initial Hamiltonian~\cite{Wen-Hao2025}:
\begin{align}
    \hat{H}_\text{eff}
    &= E_\text{ref}
    + \sum_{\ell, m\neq 0,0} E_\ell \hat b_{\ell m}^\dag \hat b_{\ell m}
    .
\end{align}
Here, the ground state energy, $E_\text{ref}$,
and the bosonic collective excitations, created by $\hat b^\dag_{\ell m}$ with energy $E_\ell$, are defined as
\begin{subequations}
\begin{align}
    E_\text{ref}
    &\equiv -\frac{1}{2} \lambda_0 4\pi \rho_0^2
    -\frac{1}{2} \sum_{\ell,m \neq 0,0} \left( \varepsilon_\ell^0 + \lambda_0 \rho_0 - E_\ell \right)
    ,
    \\
    \hat b_{\ell m}^\dag &\equiv U_{\ell} \hat{a}^{\dag}_{\ell m} - (-1)^{m} V_{\ell} \hat{a}_{\ell,-m}
    \label{eq:bogo_QP},
    \\
    U_{\ell} &\equiv i \sqrt{\left(\varepsilon_\ell^0 + \lambda_0 \rho_0 + E_\ell\right) / 2 E_\ell}
    ,
    \\
    V_{\ell} &\equiv -i  \sqrt{\left(\varepsilon_\ell^0 + \lambda_0 \rho_0 - E_\ell\right) / 2 E_\ell}
    ,
    \\
    E_\ell &\equiv
    \sqrt{\varepsilon_\ell^0 (\varepsilon_\ell^0 + 2 \lambda_0 \rho_0)}
    .
 \end{align}
\end{subequations}
Note that the collective excitation dispersion relation takes on the standard Bogoliubov form~\cite{Wen-Hao2025}.

To facilitate the computation of thermal expectation values, it is useful to write the mode expansions in terms of the Bogoliubov quasiparticle creation and annihilation operators, $\hat{b}^{\dag}_{\ell m}$ and $\hat{b}_{\ell m}$ respectively.
For the density and phase fluctuation mode expansions, Eq.~(\ref{eq:density_phase_spherical_harmonic_expansion}), we insert the inverse of Eq.~(\ref{eq:bogo_QP}), $\hat{a}_{\ell m} = U_{\ell} \hat b_{\ell m} + (-1)^{m} V_{\ell}^* \hat b_{\ell,-m}^\dag$. This yields Eq.~(\ref{modeexpansion}) in the main text where the Bogoliubov quasiparticle operator coefficients are related to the mode fluctuation operator coefficients by:
\begin{subequations}
\begin{align}
    \Pi_{\ell}(t) &\equiv U_{\ell} P_{\ell}(t) + V_{\ell} P_{\ell}^*(t)
    ,
    \\
    \Phi_{\ell}(t) &\equiv U_{\ell} Q_{\ell}(t) + V_{\ell} Q_{\ell}^*(t)
    .
\end{align}
\end{subequations}
Both sets of coefficients obey the same EOM and Wronskian condition. However, their initial conditions differ, with $\Pi_{\ell}(0) = i \sqrt{\rho_0 \varepsilon_{\ell}^0/E_{\ell}}$ and $\Phi_{\ell}(0) = \frac{1}{2} \sqrt{E_{\ell}/\rho_0 \varepsilon_{\ell}^0}$.

\section{Technical details}
\label{appendix:realisticparameter}

In this Appendix we include some 
technical details omitted from the main text, including an estimate of some system parameters, the full formula for the particle production amplitude, and show how
the particle production amplitude depends
on the quantum number $\ell$ and bubble
trap radius $R$. 

\subsection{Parameter values}
We can determine the numerical values of the parameters $a_\ell$ and $b_\ell$ in Eq.~(\ref{eqn:aandb}) directly from experimental setups. In our numerical calculations, we primarily adopt the parameters from the experimental setup of Jia et al.~\cite{Jia2022}, which utilizes a dual-species bosonic mixture consisting of a $^{87}\mathrm{Rb}$ core surrounded by a \(^{23}\mathrm{Na}\) gas shell. For the atom mass, we use the sodium atomic mass: \(M=3.82\times 10^{-26}\,\mathrm{kg}\).
The dimensions of this outer bubble shell are estimated to be $R = 10~\mu\text{m}$ in radius and $\Delta R \simeq 0.4~\mu\text{m}$ in thickness, containing $N \simeq 1.3 \times 10^5$ atoms with an initial $s$-wave scattering length of $a_s = 54.6\,a_0$.

The repulsive coupling strength $\lambda_0$ is related to the 3D interaction strength $U_0=4\pi\hbar^2 a_s / M$ via the geometric volume of the shell:
\begin{equation}
    \lambda_0=\frac{1}{\sqrt{2\pi}}
    \frac{U_0}{R^2 \Delta R}.
\end{equation}
The $s$-wave scattering length $a_s$ is widely tunable via Feshbach resonances~\cite{Chinetal2010}. Here, we assume the scattering length can be tuned up by an order of magnitude, ranging from $\sim 54.6\,a_0$ to $\sim 500\,a_0$.

To model this variation, we define the time-dependent coupling strength as $\lambda(t)=\lambda_0 e^{t/\tau}$, where the characteristic time scale $\tau$ determines the duration of the quench. In our simulations, we set $\tau=10\,\mathrm{ms}$ so that the coupling strength increases by a factor of $e^2$ over a typical experimental window of $20\,\mathrm{ms}$. The corresponding angular density, $\rho_0 = N / 4\pi$ (neglecting condensate depletion), represents the number of atoms per unit solid angle and remains independent of the bubble radius.

For comparison, we also consider an alternative set of parameters appropriate for the Cold Atom Lab (CAL)~\cite{Lundblad2019,Carollo2022}.
This setup features $^{87}\mathrm{Rb}$ atoms ($M=1.445\times 10^{-25}\,\mathrm{kg}$) configured to generate a larger bubble shell with a radius of approximately $50~\mu\text{m}$, a thickness of $1\text{--}10~\mu\text{m}$, and a total population of $N \sim 10^4$ atoms. This is the set of parameter we used in Fig.~\ref{fig:setup} to make the surface plot. The parameters corresponding to the Cold Atom Lab (CAL) describe a large, thin shell with a low atom count, resulting in low visibility in the numerical simulations. Therefore, in the main text, we choose to present plots generated using the parameter set from Jia et al.~\cite{Jia2022}. 

\subsection{Particle production amplitude}
Next, we present our rather-lengthy analytical expressions for the parameters $\alpha_\ell$ and
$\beta_\ell$, suppressing the $\ell$ subscript for
clarity, and adding the subscript $i$ to indicate
this is for the increasing-coupling case (quench up):
\begin{widetext}
\bea
&&\hspace{-0.25cm}\alpha_i(t_f) 
= \frac{i {\rm e}^{it_f \sqrt{b+a{\rm e}^{t_f}}}}{D}\Big[
  \Big(\sqrt{b+a} J_-(2\sqrt{a}) - i\sqrt{a}  J_-'(2\sqrt{a})\Big)
  \Big(\sqrt{b+a{\rm e}^{t_f}} J_+(2\sqrt{a{\rm e}^{t_f}}) + i\sqrt{a{\rm e}^{t_f}}
  J_+'(2\sqrt{a{\rm e}^{t_f}})\Big) - (-\leftrightarrow +)\Big],
  \nonumber
\\
&&\hspace{-0.25cm}\beta_i(t_f)
= \frac{i {\rm e}^{-it_f \sqrt{b+a{\rm e}^{t_f}}}}{D}\Big[
  \Big(\sqrt{b+a} J_+(2\sqrt{a}) - i\sqrt{a}  J_+'(2\sqrt{a})\Big)
  \Big(\sqrt{b+a{\rm e}^{t_f}} J_-(2\sqrt{a{\rm e}^{t_f}}) - i\sqrt{a{\rm e}^{t_f}}
  J_-'(2\sqrt{a{\rm e}^{t_f}})\Big) - (-\leftrightarrow +)\Big]
  \nonumber
\\
&& \hspace{-0.25cm} D \equiv 2\sqrt{a} (b+a)^{\frac{1}{4}} (b + a{\rm e}^{t_f})^{\frac{1}{4}}
\Big( J_+(2\sqrt{a}) J_-'(2\sqrt{a}) -  J_-(2\sqrt{a}) J_+'(2\sqrt{a}),
\Big)
\eea
\end{widetext}
where $J_\pm(x) \equiv J_{\pm 2i\sqrt{b}}(x)$ 
(with the latter denoting the Bessel function with 
imaginary index) and
$J_{\pm}'(x) \equiv \partial_x J_{\pm 2i\sqrt{b}}(x)$ and we set $\tau \to 1$.  The results
for a decreasing-coupling quench satisfy:
\bea
\alpha_d(t_f) = \alpha_i^*(-t_f),
\\
\beta_d(t_f) = \beta_i^*(-t_f).
\eea

\subsection{Dependence on $\ell$ and $R$}
Finally, we show that the 
dependence of the particle production 
amplitude on $\ell$ and $R$ takes
a relatively simple form when expressed in terms of the effective wavevector 
$k =\sqrt{\ell(\ell+1)}/R$.  To
see this, we note that,  although our final result for $\beta_\ell$ is rather complicated, in fact there is an underlying simplification. This follows from the fact that the dependence of $\beta_\ell$ on the quantum number $\ell$ and the system parameters (i.e., $R$, $\Delta R$, $\rho_0$ and $T$) enter only via the following dimensionless ratios:
\begin{eqnarray}
    \frac{\varepsilon_\ell}{2 \lambda_0 \rho_0} &=& \frac{b_\ell}{a_\ell} = \sqrt{\frac{2}{\pi}}\frac{\Delta R}{a_s  \rho_0} \ell(\ell + 1), \label{eq:curv1}\\
    \frac{\varepsilon_\ell}{(\hbar/\tau)} &=& \sqrt{b_\ell} = \frac{\hbar \tau}{2 m R^2} \ell(\ell + 1),
    \label{eq:curv2}\\
    \frac{k_{\mathrm{B}} T}{\varepsilon_\ell} &=& 
    \label{eq:curv3}
    \frac{k_{\mathrm{B}} T}{\hbar^2 / (2 m R^2)} \frac{1}{\ell(\ell + 1)}.
\end{eqnarray}
Now consider the $\ell$ and $R$ dependence of these quantities.  The last two (Eqs.~(\ref{eq:curv2}) and (\ref{eq:curv3})) explicitly depend only on these quantities in the ratio $k =\sqrt{\ell(\ell+1)}/R$, an effective wavevector.  In the first equation, if we work at fixed atom density $\rho$ (so that $\rho_0 \propto \rho R^2$), Eq.~(\ref{eq:curv1}) also only depends on $k$.  This shows, as discussed in the main text, that the particle production amplitude $\beta_\ell$ is
approximately independent of $R$ as long
as it is plotted as a function of $k$. 

\end{document}